%% file: main.tex
\documentclass[10pt, journal, letterpaper]{IEEEtran}
\IEEEoverridecommandlockouts
\usepackage{xspace}
\usepackage{threeparttable}
\usepackage{color}
\usepackage{url}
\usepackage{subcaption}
\usepackage{paralist}
\usepackage{wrapfig}
\usepackage{multirow}
\usepackage{listings}
\usepackage{booktabs}
\usepackage{makecell}
\usepackage{boxedminipage}
\usepackage{amsmath}
\usepackage{graphicx}
\usepackage{minibox}
\usepackage{algorithm}
\usepackage{comment}
\usepackage{pifont}
\usepackage{etoolbox}
\usepackage{cite}

\usepackage{amssymb}
\usepackage{color}
\usepackage{xcolor}
\usepackage{subfiles}
\usepackage[utf8]{inputenc}
\usepackage[english]{babel}
\usepackage[cachedir=.]{minted}
\usepackage{xcolor}
\usepackage{ulem}
\usepackage{adjustbox}
\usepackage{diagbox}
\usepackage{colortbl}
\usepackage{balance} 
\usepackage{multicol}
\newcounter{rowID}
\newcommand{\nextID}{\stepcounter{rowID}\therowID}

\newcommand{\ra}[1]{\renewcommand{\arraystretch}{#1}}

\newcommand{\yt}[1]{\textcolor{black}{#1}}

\newcommand{\yu}[1]{\textcolor{black}{#1}}

\input{macro.tex}
\begin{document}

\title{Unraveling Responsiveness of Chained BFT Consensus with Network Delay}

\author{
    Yining Tang$^{*}$, Mohan Yu$^{*}$, Qihang Luo, Runchao Han,
    Jianyu Niu,~\IEEEmembership{Member, ~IEEE,} \\
    Chen Feng,~\IEEEmembership{Member, ~IEEE,} 
    Yinqian Zhang,~\IEEEmembership{Member, ~IEEE} 
    
    \IEEEcompsocitemizethanks{
        \IEEEcompsocthanksitem Yining Tang is with the Research Institute, China Telecom Company Ltd., Guangzhou 510660, China. Email: tangyn7@chinatelecom.cn.

        Mohan Yu, Qihang Luo, Jianyu Niu, and Yinqian Zhang are with the Research Institute of Trustworthy Autonomous Systems and the Department of Computer Science and Engineering, Southern University of Science and Technology, Shenzhen, China.
        Email: yumohan01@gmail.com, 12110425@mail.sustech.edu.cn, niujy@sustech.edu.cn and yinqianz@acm.org.
        
        Chen Feng is with Blockchain@UBC and the School of Engineering, The University of British Columbia (Okanagan Campus), Kelowna, BC, Canada. Email: chen.feng@ubc.ca. 

        Runchao Han is at Babylon Labs. E-mail: runchao@babylonlabs.io.

        $^{*}$Authors are equally contributed. 
    }
}
\sloppy

\maketitle

\begin{abstract} 
With the advancement of blockchain technology, chained Byzantine Fault Tolerant (BFT) protocols have been increasingly adopted in practical systems, making their performance a crucial aspect of the study. In this paper, we introduce a unified framework utilizing Markov Decision Processes (MDP) to model and assess the performance of three prominent chained BFT protocols. Our framework effectively captures complex adversarial behaviors, focusing on two key performance metrics: \metricOne and \metricTwo. 
We implement the optimal attack strategies obtained from MDP analysis on an existing evaluation platform for chained BFT protocols and conduct extensive experiments under various settings to validate our theoretical results.  
Through rigorous theoretical analysis and thorough practical experiments, we provide an in-depth evaluation of chained BFT protocols under diverse attack scenarios, uncovering optimal attack strategies. Contrary to conventional belief, our findings reveal that while responsiveness can enhance performance, it is not universally beneficial across all scenarios.
This work not only deepens our understanding of chained BFT protocols, but also offers valuable insights and analytical tools that can inform the design of more robust and efficient protocols.
\end{abstract}


\section{Introduction} \label{sec:intro}
The growing popularity of decentralized applications, including global payments~\cite{nakamoto2012bitcoin}, DeFi~\cite{wood2014ethereum}, IoT~\cite{TC1, GaiNiu}, mobile crowdsensing~\cite{TongZhouIncentive}, and online gaming\footnote{Crypto Kitties: \url{https://www.cryptokitties.co/}}, has revitalized interest in Byzantine Fault Tolerant (BFT) consensus~\cite{castro1999practical}. 
Recently, a family of chained BFT protocols utilizing the chaining structure of blockchains has attracted extensive attention for their ability to support large-scale decentralized applications. 
Notable examples include Tendermint~\cite{Buchman2016TendermintBF}, Casper FFG~\cite{casper}, HotStuff~\cite{hotstuffPODC},  Streamlet~\cite{chan2020streamlet}, Fast-HotStuff~\cite{fastHotStuff}, and HotStuff-2~\cite{malkhi2023HotStuff}. 
These protocols have been used in tens of blockchains, both 
permissioned (\eg, XuperChain\footnote{XuperChain Platform: \url{https://github.com/xuperchain/xuperchain}} and 
Hyperchain) and permissionless (\eg, Ethereum 2.0\cite{schwarz2022three}, Aptos\footnote{Aptos: \url{https://aptoslabs.com}}, Cypherium, Flow\cite{hentschel2002flow}, Zilliqa 2.0, and DeSo\footnote{Deso: \url{https://revolution.deso.com/}}).

Tendermint~\cite{Buchman2016TendermintBF} and Casper FFG~\cite{casper} are among the first generations of chained BFT protocols that utilize chain structure and pipelining techniques to realize linear message complexity and improve system efficiency, respectively. 
However, they cannot guarantee responsiveness, by which a designated leader can drive nodes to reach a consensus in time depending only on the actual message delays (denoted as $\delta$), independent of any known upper bound delays (denoted as $\Delta$). 
The upper bound delay represents the worst-case time for messages to propagate over the network, typically set to be more than an order of magnitude larger than the actual delay. 
Thus, achieving responsiveness is a hallmark of practical BFT protocols, as claimed in~\cite{hotstuffPODC}. 
Due to its importance, subsequent chained BFT protocols such as chained HotStuff (\CHS) and Fast-HotStuff (\FHS) adopt different designs to achieve it. 
Specifically, \CHS extends two-phase message exchanges to three-phase exchanges, whereas \FHS requires blocks to include QCs (\ie, a set of collected messages from more than two-thirds of nodes to denote the highest block they have seen). 
Despite the introduced protocol overhead, responsiveness is considered a crucial property in BFT protocols, as it can significantly reduce latency under optimistic conditions, guiding all following designs~\cite{sui2022marlin, giridharan2023beegees, malkhi2023HotStuff}.

Unfortunately, the responsiveness belief behind chained BFT protocols has not been carefully examined.
The prior benchmarks~\cite{dinh2017blockbench, shapiro2020performance,amiri2024bedrock,gramoli2023diablo} try to evaluate the performance (\ie, throughput and latency) of chained BFT protocols under various experimental settings. 
However, they mainly evaluate performance in the ideal cases, \ie, all nodes follow the protocol. 
Studies~\cite{niu2021performance, gai2021dissecting} have shown that in chained HotStuff, an adversary can strategically fork uncommitted blocks from honest leaders to weaken the system performance.
Since chained BFT protocols are usually deployed among distrusting nodes, it is important to analyze or evaluate the impact of responsiveness on performance under attacks.  
More importantly, to fairly compare chained BFT with/without responsiveness, we should determine their worst performance and the associated optimal adversarial attack strategies. To our knowledge, none of the existing studies can achieve this. 

\begin{figure}[t]
    \flushleft
    \includegraphics[width=1\linewidth]{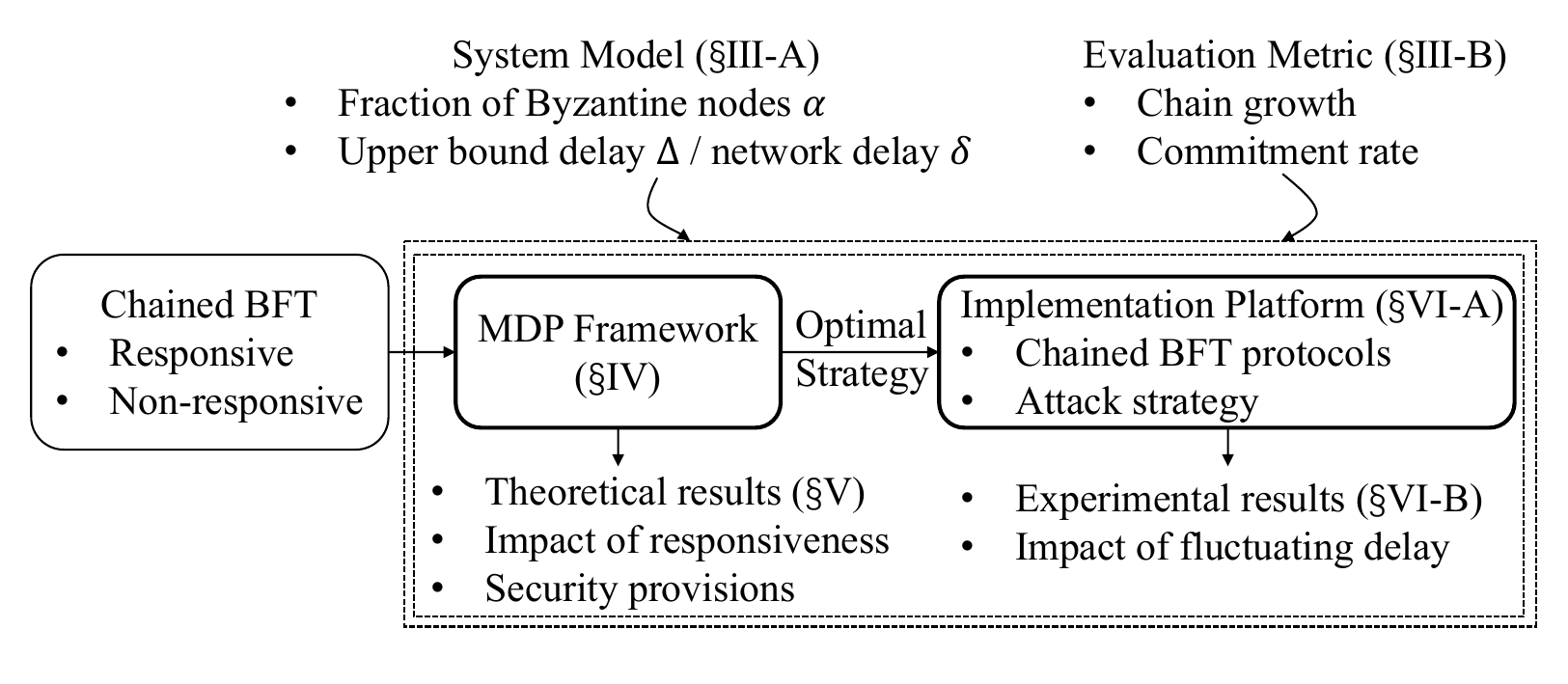}
    \caption{An overview of the proposed framework.}
    \label{fig:overview}
    \vspace{-6mm}
\end{figure}

In this paper, we propose a comprehensive framework to analyze the impact of responsiveness on system performance. 
With the framework, we can analyze the performance of existing chained BFT protocols, \ie, chained HotStuff (CHS), Two-Chain HotStuff (2CHS), and Fast-HotStuff (FHS) as well as other chained BFT protocols, under attacks.  
Instead of using traditional throughput and latency metrics~\cite{dinh2017blockbench, shapiro2020performance,amiri2024bedrock,gramoli2023diablo}, we propose two customized performance metrics: \textit{chain growth}, which is the rate at which honest blocks are added to the main chain, reflecting the system's capacity to process transactions, and \textit{commitment rate}, which indicates the frequency of block commitment events on the main chain, showing the speed of transaction confirmation.
Both metrics focus on the blocks instead of transactions (usually used to measure traditional throughput and latency metrics), which enables us to ignore unnecessary details like transaction batch size and block size. 
Thus, these two metrics not only reflect performance under attack scenarios but also are easier to trace in Markov Decision Processes (MDP) modeling. 

Our framework consists of two key components:
(i) an MDP-based analytical model and (ii) an evaluation platform, as shown in Fig.~\ref{fig:overview}. 
The first component enables us to explore optimal attack strategies in chained BFT protocols by considering factors such as the fraction of Byzantine nodes, the impact of network delay, and various consensus rules. MDP provides a mathematical framework for simulating and analyzing decision processes in random environments, making it highly suitable for quantifying the impact of arbitrary behaviors of random-selected adversarial leaders on protocol performance.

With this MDP framework, we can fairly compare the performance of responsive and non-responsive chained BFT protocols under diverse attack scenarios.  
Although MDP has been used to analyze the performance and security of the Nakamoto-style consensus~\cite{sapirshtein2017optimal, gervais2016security, NiuBitcoinNG}, its application to chained BFT protocols remains challenging due to fundamental design differences. 
{Unlike Nakamoto’s PoW (fork resolution and probabilistic guarantee), chained BFT relies on leader-driven, quorum-based deterministic commits. Thus, we redefined the state space to include BFT-specific attributes (leader rotation, quorum progress, pending blocks) absent in Nakamoto-focused models. The accommodation demands an in-depth comprehension of both MDP tools and the details of chained BFT protocols.}
As one of our contributions, we properly simplify the system state and constrain adversary actions to facilitate MDP modeling of various chained BFT protocols.

The second component enables us to implement the optimal attack strategies identified by the MDP modeling on some implementation platforms of chained BFT protocols. In this work, we adopt the open-sourced Bamboo platform, which supports CHS, FHS, and Streamlet~\cite{chan2020streamlet}.  
Although Bamboo also supports performance evaluation under attacks, these strategies are straightforward and so cannot provide a fair comparison between chained BFT protocols.   
Our extension to Bamboo not only can validate the theoretical results from MDP modeling, but also can evaluate performance with consideration of more factors of practical systems such as fluctuating network delay. 

Our work provides a way to holistically compare the impact of responsiveness on the system performance. For instance, when the fraction of Byzantine nodes is small, responsive protocols can achieve better chain growth and commitment rate.  Our key findings are summarized as follows:
\begin{packeditemize}
    \item \textit{Finding 1}: Across all three protocols, both \metricOne and \metricTwo degrade significantly under attacks, with the impact worsening as the fraction of Byzantine nodes increases. \yu{While responsive protocols can achieve better performance at low Byzantine fractions, their performance advantage diminishes as adversarial influence grows.}
    

    \item \textit{Finding 2}: \yu{While responsive designs can improve performance under dynamic conditions, adding mechanisms to achieve responsiveness does not always yield benefits. The design must balance the cost of these mechanisms with protocol-specific requirements under adversarial behaviors.}


    \item \textit{Finding 3}: Our experiments validate the framework’s effectiveness, showing that \metricOne and \metricTwo remain robust even under adverse and fluctuating network conditions.
\end{packeditemize}

\bheading{Our contributions.} The contributions of this paper are listed below. 

\begin{packeditemize}
   \item 
   We propose an evaluation framework with two metrics, \metricOne and \metricTwo, to systematically analyze the performance of chained BFT protocols.

   \item 
   We model and evaluate three representative chained BFT protocols under different fractions of Byzantine nodes based on the proposed framework. 
   We obtain the optimal theoretical results and attack strategies. Through an analysis of the results under different fractions, we find the impact of responsiveness on protocol performance

   \item We conduct extensive empirical experiments to validate the theoretical framework in real-world settings, examining the impact of network delay fluctuations on protocol performance. The integration of theoretical analysis and empirical validation enhances the reliability of our findings.
   
\end{packeditemize}

\vspace{1mm} \bheading{Roadmap.} 
\secref{sec:background} introduces background and related work on chained BFT protocols. 
\secref{sec:model} provides the system model and metrics.
\secref{sec:MDPmodel} presents an MDP model of chained BFT protocols.
The theoretical results of the MDP model and experimental results are provided in \secref{sec:MDP-eval} and \secref{sec:experiments}, respectively. 
The extension of Streamlet is given in \secref{sec:beyond}. 
We introduce extensions of the framework in \secref{sec:discussion} and conclude the paper in \secref{sec:conclusion}.

\section{Background and Related Work} \label{sec:background}
We first provide the background of chained BFT protocols and the associated performance issues. 
Then, we review prior studies on the performance analysis of chained BFT protocols.

\subsection{Chained BFT Consensus} \label{subsec:CBFT}
\begin{figure}
    \centering
    \includegraphics[width=1\linewidth]{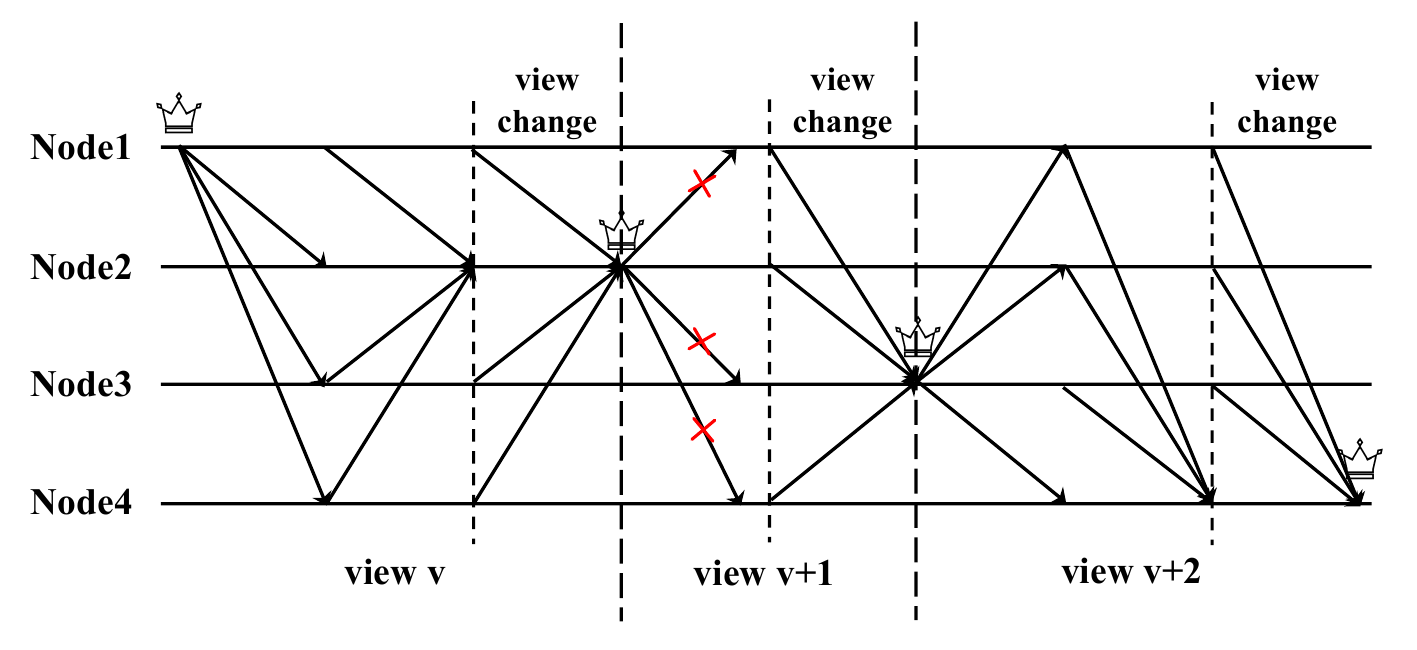}
    \caption{The consensus process of chained BFT protocols.}
    \label{fig:paradigm}
    \vspace{-3mm}
\end{figure}

Chained BFT consensus represents a family of BFT protocols that leverage the chain structure for better performance and scalability~\cite{fastHotStuff,hotstuffPODC,chan2020streamlet}. Chained BFT protocols run in views, in which a designated replica (called the leader) coordinates with others to vote for its block, as shown in \figref{fig:paradigm}. 
Specifically, each view can be further divided into two stages: the leader-based stage and the view-change stage. 
In the former, the leader proposes a new block to extend previous blocks, \ie, forming a chain, according to the proposing rule. 
Then, other nodes append the first valid block from the leader to their chain and update the local state of the locked block.
Later, they vote for the block by the voting rule. Once a block has enough votes (from more than $2/3$ of the nodes), it forms a Quorum Certificate (QC) and is certified. 
Node follows the committing rule to check whether an uncommitted block in the chain can be committed. Note that once a block is committed, its uncommitted ancestor blocks are also committed.

In the view-change stage, nodes safely wedge to the next round if the leader is faulty or the proposed block's quorum certificate is not formed before the timeout. This stage is also referred to as Pacemaker~\cite{hotstuffPODC}. 
It is crucial for the \textit{liveness} property, by which honest clients' transactions are eventually included in committed blocks.  
Meanwhile, chained BFT protocols also need to ensure the \textit{safety property}, by which honest nodes accept the same committed chain of blocks, referred to as the \textit{main chain}---a key concept that will be used in our analysis.
In other words, a block is included in the main chain once it is committed.  

\bheading{Responsiveness.} The responsiveness property ensures that a designated leader can drive nodes to reach consensus solely at the speed determined by actual network delays ($\delta$), without relying on any predefined upper bound on delays ($\Delta$).
Specifically, the upper bound delay $\Delta$ guarantees message delivery between any two honest nodes after Global Stabilization Time (GST), concerned with system liveness, whereas the network delay $\delta$ directly affects the system performance. Thus, responsiveness foregoes a hallmark of practical BFT protocols, as claimed in~\cite{hotstuffPODC}.  

Due to its importance, many chained BFT protocols adopt different designs to achieve responsiveness. 
Specifically, Two-Chain HotStuff (\TCHS) formulated from Tendermint~\cite{Buchman2016TendermintBF} and Casper FFG~\cite{casper} is the original chained BFT protocol without responsiveness.
{In contrast, chained HotStuff (\CHS) achieves responsiveness by extending the two-phase exchanges to three-phase exchanges.} Fast-HotStuff (\FHS) introduces the latest QC proof and achieves responsiveness within two-phase exchanges.  
\FHS also optimizes the protocol to skip the view change in the happy path. 
As said previously, achieving responsiveness or not directly affects system performance, further determining its practicality. \yu{Thus, exploring how mechanisms for achieving responsiveness affect protocol performance is an important question.}
 

\bheading{Performance under attacks.} The performance of BFT protocols is usually measured in terms of throughput and latency. Throughput refers to the number of transactions that can be processed per unit of time, while latency measures the time taken to reach consensus on a transaction. Applying BFT protocols to blockchain introduces the concept of blocks, leading to a greater focus on honest blocks.

The performance of chained BFT protocols can be affected by Byzantine nodes. 
An \yu{adversarial leader} may intentionally cause delays in the protocol. For example, it can put off the decision-making or vote-collecting process, leading to slow consensus progress before triggering a timeout. It can also remain silent~\cite{gai2021dissecting} or propose invalid blocks, causing the entire consensus process to stall until a timeout occurs. 
The \yu{adversarial leader} can also exclude honest blocks from the chain or disrupt the committing rule of blocks~\cite{niu2021performance}, causing the transactions in the block to be delayed. 
Therefore, it is essential to identify potential vulnerabilities and then measure their impact on performance. 

\subsection{Related Work}
We first introduce some state-of-the-art chained BFT protocols. Then, we review existing studies to analyze the performance of chained BFT protocols, as well as MDP modeling of blockchain consensus protocols. \tabref{tab:relatedWork} lists existing benchmarks and evaluation frameworks of chained BFT protocols.

\begin{table}[t]
\caption{Summary of prior frameworks to evaluate the performance of chained BFT protocols under attacks.} 
\label{tab:relatedWork}
\resizebox{\linewidth}{!}{
\centering
\ra{1.05}
\begin{tabular}{@{}cccc@{}}
\toprule[1pt]
\multirow{2}{*}{Framework} & \multicolumn{2}{c}{Methodology} & \multirow{2}{*}{Optimal Strategy} \\ 
 & Experimental & Theoretical & \\
\midrule[1pt]
~\cite{gai2021dissecting,dinh2017blockbench, shapiro2020performance,amiri2024bedrock,gramoli2023diablo,cohen2022aware} & \ding{51} & \ding{55} & \ding{55}\\
~\cite{niu2021performance} & \ding{55} & \ding{51} & \ding{55} \\
\midrule
This work & \ding{51} & \ding{51} & \ding{51} \\
\midrule[1pt]
\end{tabular}}
\vspace{-6mm}
\end{table}

\bheading{Chained BFT protocols.}\label{subsec:cbftProtocols} 
Beyond \CHS, \TCHS, and \FHS mentioned above, there are other variants of chained BFT protocols. For instance, Streamlet~\cite{chan2020streamlet} simplifies protocol design with different consensus rules; Jolteon~\cite{gelashvili2022jolteon} and DiemBFT\_v4~\cite{team2021diembft} employ a quadratic view change mechanism, enabling two-phase block commitment under steady state; 
Wendy~\cite{giridharan2021no} leverages no-commit proofs to achieve optimal latency and linear message complexity; Marlin~\cite{sui2022marlin} introduces a ranking system and virtual blocks to counter forking attacks and reduce latency and bandwidth consumption.
Despite differences in implementation details, these variants maintain the consensus paradigm of chained BFT. Thus, our framework may be extended to study these protocols.

\bheading{Performance analysis of chained BFT protocols.} 
Prior benchmarks~\cite{dinh2017blockbench, shapiro2020performance,amiri2024bedrock,gramoli2023diablo,gai2021dissecting} focus on evaluating the throughput and latency metrics of various BFT protocols under different experimental settings.
However, most of them do not consider the impact of attacks. 
Besides, the accuracy of evaluation results is affected by many factors such as implementation and deployment environment.
Few studies analyze the performance of \CHS protocols under attacks. Several work~\cite{niu2021performance, niu2025chained, tang2025leader} mention delay attack and forking attack, and provide theoretical analysis. Cohen et al.~\cite{cohen2022aware} design an attack and quantify its impact on throughput and latency through experiments. However, these studies only focus on a single protocol and cannot be directly adapted to analyze other protocols.

\bheading{MDP modeling of blockchain consensus.}
There have been extensive studies~\cite{sapirshtein2017optimal, gervais2016security, NiuBitcoinNG, feng2019selfish, zhang2019lay} that apply 
MDP to evaluate the security and/or performance of blockchain consensus protocols.
However, they focus on Nakamoto-style consensus, e.g., Bitcoin~\cite{nakamoto2012bitcoin} or Ethereum~\cite{wood2014ethereum}.
Their models cannot be directly applied to analyze chained BFT protocols that follow significantly different design principles from Nakamoto-style consensus. {Unlike Nakamoto-style consensus, which lacks leader nodes and relies on solving mathematical puzzles to propose blocks, chained BFT protocols provide deterministic finality \cite{hotstuffPODC}: once a block is committed, it cannot be reverted. These fundamental differences necessitate the development of new models specifically for chained BFT protocols.}

\section{System Model and Metrics} \label{sec:model}
\subsection{System Model}
We consider a system with $f$ Byzantine nodes among $n$ ($n\geq3f+1$) nodes by following existing chained BFT protocols~\cite{casper, hotstuffPODC, fastHotStuff, malkhi2023HotStuff, NiuEBFT}. The Byzantine nodes can behave arbitrarily, while honest nodes always follow the consensus rules.
We define $\alpha\!=\!f/n$ as the fraction of Byzantine nodes. \yu{In our framework, we model leader selection as an i.i.d. Bernoulli process, where each view elects an adversarial leader with probability $\alpha$. 
While production protocols typically use deterministic rotation, this stochastic abstraction keeps the MDP tractable and captures the steady-state impact of adversarial leaders.} \yt{In production protocols with deterministic rotation, a sequence-aware adversary may exploit the predictable leader schedule to plan attacks more effectively. Deterministic schedules may introduce additional sequence effects beyond scope of this abstraction. }
\yu{Thus, our stochastic model should be interpreted as a tractable steady-state baseline under a fixed adversarial fraction.} We consider the worst-case situation where a single adversary controls all Byzantine nodes. 
The adversary attempts to degrade system performance (\ie, minimizing the chain growth and commitment rate introduced shortly) with optimal strategies.


\bheading{Network model.} We assume a partially synchronous model by following previous work~\cite{niu2021performance, amiri2024bedrock, gramoli2023diablo, gai2021dissecting, TC3}, ensuring all honest nodes are in sync with the protocol's state after Global Stabilization Time (GST). 
In particular, after GST, a message is guaranteed to be delivered within a known and bounded delay $\Delta$.
This paper focuses on analyzing chained BFT protocols during network synchrony (\ie, after GST) as prior analysis or evaluation works~\cite{niu2021performance, amiri2024bedrock, gramoli2023diablo, gai2021dissecting}. We argue that during network asynchrony~\cite{TC2} (\ie, before GST), the adversary can cause more damage to the performance by utilizing network partitions. {For example, an honest block may not be received by all honest nodes during asynchrony, which gives the adversary a chance to propose a block to override it. However, our analysis during network synchrony already reveals many insights about the design. Therefore, for simplicity and consistency with prior work, we focus on the synchronous setting and leave the asynchronous case to future work.}

To show the impact of responsiveness, we assume the actual network delay between two nodes is $\delta$ by following prior work~\cite{momose2020hybrid,abraham2020sync, hotstuffPODC}. 
This delay varies depending on the real-time network conditions. 
We assume a fixed $\delta$ in the analysis, 
while considering varied $\delta$ in our experiments.  We use $k$ to denote the ratio between the maximum bound delay $\Delta$ and the actual network delay $\delta$, \ie, $k = \Delta / \delta$. \yu{For simplicity, the theoretical analysis treats $\delta$ as fixed, while our experiments evaluate fluctuating delays. 
The model can be extended to stochastic or non-uniform delay settings by replacing the fixed $\delta$ with per-view delays sampled from a delay distribution, which may change exact metric values but not the main trends.}

\subsection{Evaluation Metrics} \label{subsec:metrics}
To evaluate the performance of chained BFT protocols, we introduce two metrics, \ie, \metricOne and \metricTwo, as introduced below. 

\subsubsection{\metricOneUp}
The \metricOne $G(\alpha)$ is defined as the rate of honest blocks appended to the main chain, given that the adversary corrupts a fraction $\alpha$ of the total nodes. \yu{In this paper, we refer to such an event as block inclusion.} It indicates the impact of the adversary on system efficiency. Specifically, the adversary can strategically minimize \metricOne to reduce the system capacity for processing clients' transactions, leading to significant service delays or system congestion. {Since our analysis assumes a saturated workload, excluding extreme worst cases (\eg, empty honest blocks), the system is continuously processing transactions. Under this assumption, any reduction in chain growth directly corresponds to a reduction in block inclusion.} Thus, we capture this by minimizing \metricOne calculated in $m$ views. We use ${B_h}_{i}$ and ${T}_{i}$ to denote the number of honest blocks added to the main chain in the $i$-th view and time units consumed in the $i$-th view, respectively. We have:
\begin{equation}
    G(\alpha) = \lim_{m \rightarrow \infty}{\frac{\sum_{i=1}^{m} {B_h}_{i}}{\sum_{i=1}^{m} {T}_{i}}}
\end{equation}

This metric considers only honest blocks, since adversarial leaders can selectively exclude honest clients' transactions. {Moreover, all transactions in honest blocks are first verified by honest nodes before inclusion, as in most blockchain systems.} Thus, {this metric} can reflect the \textit{effective} capacity of processing transactions. 
To lower \metricOne, the adversary can either prevent honest blocks from being included in the main chain or delay block generation.
This metric is first proposed in~\cite{niu2021performance, gai2021dissecting} to evaluate the performance of CHS.

\subsubsection{\metricTwoUp}
The \metricTwo $R(\alpha)$ is defined as the rate of block commitment event at the main chain, given the fraction of adversarial nodes $\alpha$. 
This metric captures system stability to continuously commit blocks and then confirm clients' transactions. 
The adversary may target to minimize this metric to delay transaction confirmation, increasing transaction delay to affect time-sensitive applications.   
We denote whether blocks are committed in the $i$-th view by $C_i$, which is assigned the value of 1 if a commitment occurs in the $i$-th view, and 0 otherwise. Here, in chained BFT protocols, when a block is committed, all its uncommitted ancestor blocks are also committed. Thus, in a commitment event, several blocks may be committed. We have:
\begin{equation}
    R(\alpha) = \lim_{m \rightarrow \infty}{\frac{\sum_{i=1}^{m} {C}_{i}}{\sum_{i=1}^{m} {T}_{i}}}
\end{equation}

In chained BFT protocols, a block has to satisfy the commitment conditions to be committed (see Sec.\ref{subsec:CBFT}). 
Thus, the adversary deviates from the protocol to ruin the conditions. 
Note that this metric can also serve as an indicator of \textit{liveness}. 
For example, if $R(\alpha)$ is $0$ for a chained BFT protocol, it means no transactions can be committed, \ie, no liveness.

\iheading{Discussion.} Compared to commonly used throughput and latency metrics~\cite{dinh2017blockbench, amiri2024bedrock, TC4, gramoli2023diablo,niu2022crystal, TC5, TC6}, these two metrics reflect performance under attacks more directly and are easier to trace in MDP modeling (introduced shortly).
First, given the maximum batch size of transactions included in a block, \metricOne directly reflects the maximum capacity of transactions included in honest nodes' blocks, \ie, the \textit{effective} throughput. Meanwhile, \metricTwo measures the frequency of block commitment, which is directly related to the time it takes for transactions to be confirmed, i.e., latency. Second, two metrics focus on the block level, whereas throughput and latency reflect system performance at the transaction level. {Focusing on block-level modeling helps simplify the analysis, which is also consistent with common practice. Specifically,} ignoring unnecessary details makes them more traceable for analyzing strategies and state transitions in MDP modeling.

Overall, these two metrics simplify MDP modeling while maintaining a direct connection to user-perceived performance. With a fixed block size and saturated workload, \metricOne maps to effective throughput, while \metricTwo reflects the average interval between commitment events and approximates confirmation latency. \yu{For example, if $\delta=100$ ms and each honest block contains 1,000 valid transactions, then $G(\alpha)=0.3$ gives an effective throughput of $0.3 \times 1000 / 0.1 = 3000$ TPS. 
Similarly, reducing $R(\alpha)$ from 0.2 to 0.05 increases the average commitment interval from $0.1/0.2=0.5$ seconds to $0.1/0.05=2$ seconds.
} \yu{We summarize the key notations used in the system model, evaluation metrics, and MDP modeling in \tabref{tab:notation}.}

\begin{table}[t]
\centering
\caption{\yu{Key Notations}}
\label{tab:notation}
\small
\renewcommand{\arraystretch}{1.05}
\scalebox{0.79}{
\begin{tabular}{ll|ll}
\toprule
Notation & Description & Notation & Description \\
\midrule
$s$ & MDP state
& $\delta$ & Actual network delay \\

$n$ & Number of nodes  
& $\Delta$ & Maximum network delay \\

$k$ & Delay ratio, $\Delta/\delta$ 
& $f$ & Number of Byzantine nodes  \\

$\alpha$ & Byzantine fraction, $f/n$
&  $T_i$ & Time cost in view $i$ \\

$G(\alpha)$ & Chain growth 
&  $B_{h_i}$ & Honest blocks in view $i$  \\

$R(\alpha)$ & Commitment rate
& $C_i$ & Commitment indicator in view $i$ \\

$L$ & Leader type, $H$ or $A$
& $l_a$ & Uncommitted adversarial blocks \\

$cS$ & Consecutive-block state 
&  $l_h$ & Uncommitted honest blocks \\
\bottomrule
\end{tabular}
}
\end{table}


\section{MDP Modeling} \label{sec:MDPmodel}
In this section, we use MDP to model three chained BFT protocols, \ie, \TCHS, \CHS, \FHS,  to show the impact of responsiveness on \metricOne and \metricTwo. We choose to model these three protocols as i) they achieve distinct responsiveness guarantees, and evaluating them will reveal insights on responsiveness, and ii) they are classic protocols that inspire subsequent protocols and have been successfully integrated into practical blockchain platforms~\cite{kwon2016cosmos}. A detailed description of these three protocols is provided in \ssecref{subsec:CBFT}. 

\subsection{Overview}
The Markov Decision Process (MDP) is a powerful mathematical framework for formalizing the decision-making process of an agent within a given environment. MDP captures the randomness in the agent's decision-making and correlates the rewards it receives with the current state of the environment~\cite{sigaud2013markov, puterman2014markov}. 
MDP provides a way of representing actions taken in a sequence of states, where each state transition is associated with reward allocations. It enables us to find optimal attack strategies that may uncover vulnerabilities within each protocol. Furthermore, we can understand and defend against these found attacks. The framework {models the consensus process and potential attacks across several chained BFT protocol designs, also enabling a comparison between them.

The MDP modeling includes four fundamental components: $\langle S, A, P, R \rangle$. 
In the context of chained BFT protocols, 
$S$ is a set of all possible states that denote the current status of the chaining blocks and affect the decision-making process. $A$ is the set of all possible actions that can be taken by the agent, \ie, the adversary. The actions alter the state of the chaining blocks and determine the corresponding rewards. 
$P$ corresponds to the transition probability function, which represents the likelihood of moving from one state to another based on the action. 
$R$ reflects the benefits or penalties incurred by the system as a result of the transition. The reward is not the incentives obtained by nodes participating in consensus, but rather the reward of terms in calculating the metrics.
The reward is essential in guiding the policy search toward actions that yield the most favorable outcomes.

{Building on the linearization and binary search construction from Sapirshtein et al. [21], we make targeted adaptations to fit chained BFT’s characteristics, addressing the mismatch between PoW-oriented MDP designs and BFT’s deterministic logic.}
Specifically, modeling chained BFT protocols (\ie, state and action space) presents several challenges. First, it demands an accurate abstraction of the protocol, stripping away irrelevant details that do not influence the two metrics while accounting for all possible malicious behaviors. Second, the size of the state and action spaces must be carefully managed to ensure the model remains computationally solvable. Finally, the MDP model must be flexible enough to accommodate multiple chained BFT protocols, further complicating its design. {To this end, we abstract and simplify the protocols, constraining the adversary’s actions to construct tractable MDP models for different chained BFT protocols, while preserving the accuracy of the results.} 

\subsection{\yu{Generic MDP Model}}
We first present the common MDP model used for chained BFT protocols.
The model is represented as $\langle S,A,P,R\rangle$, where $S$ is the
state space, $A$ is the adversary's action space, $P$ is the transition
probability function, and $R$ is the reward function used to calculate the
performance metrics.

\bheading{State space.}
We use a four-tuple state
\[
s=(cS,l_a,l_h,L).
\]
Here, $cS$ represents the consecutive block structure that affects
commitment, $l_a$ represents the number of uncommitted adversarial blocks,
$l_h$ represents the number of uncommitted honest blocks, and
$L\in\{H,A\}$ represents whether the current leader is honest or
adversarial. The exact value ranges of these variables are specified
separately for each protocol.

\bheading{Action space.} The adversary can deviate from the prescribed consensus rules to maximize their benefits. The following actions defined in the MDP framework are available to the adversary.
\begin{packeditemize}
    \item \textbf{\texttt{Adopt}}. The adversary adopts all available honest blocks that are not locked and discards the hidden adversarial blocks. {This action represents the adversary's choice to halt any attack and proceed along the honest main chain.} If the adversary is the leader, it will propose a new block that extends the last adopted block. If the adversary is not the leader, it waits for an honest leader to propose a new block.

    \item \textbf{\texttt{Wait}}. If the adversary is elected as the leader, it will create a forking block or extend its prior forking block, to override honest blocks. Otherwise, it will wait for the honest leader to propose a new block.

    \item \textbf{\texttt{Release}}. This action is only available when there is a hidden adversarial block. After releasing the block at the beginning of the view, the leader will propose a new block to extend it. If the hidden block is a forking block, the non-locked honest blocks will be overridden.

    \item \textbf{\texttt{Silent}}. The adversary remains inactive in this view. If the adversary is the leader, it either proposes no block or an invalid one. If the leader is honest, the silence of the adversary does not affect the protocol, and honest nodes continue to vote for the block proposed in this view.
\end{packeditemize}

{In chained BFT protocols, the adversary can also propose equivocating blocks or selectively drop messages during a view. However, these behaviors weaken the adversary's power to win forks, making attempts to reduce chain quality and commitment rate suboptimal. Therefore, we do not consider these attacks.}

\bheading{Transition probability.}
After each action, the next leader can be adversarial with probability
$\alpha$, or honest with probability $1-\alpha$, i.e.,
\[
\Pr[L'=A]=\alpha,\qquad \Pr[L'=H]=1-\alpha .
\]
Therefore, the transition probability $P$ is determined by the current
state, the adversary's action, and the leader type of the next view.

\bheading{Reward function.}
For each transition, the reward records the variables used to calculate
chain growth and commitment rate. Specifically,
\[
R(s,a,s')=(B_h,C,T),
\]
where $B_h$ denotes the number of honest blocks added to the main chain,
$C$ denotes the number of commitment events, and $T$ denotes the time
consumed in the transition. The detailed calculation of these rewards is
then instantiated for each protocol.

Based on this common model, we next instantiate the state ranges, state
transitions, and reward allocation for CHS, 2CHS, and FHS.

\subsection{Modeling Chained BFT Protocols} \label{subsec:CBFTmodel}
\yu{We now instantiate the generic MDP model for three chained BFT protocols.} Due to space constraints, we take \CHS as an example to illustrate the modeling process and only describe the differences in the modeling of other protocols.

\subsubsection{\CHS} CHS is the first chained BFT protocol that introduces responsiveness to accelerate the consensus process. Specifically, it introduces one extra phase to the consensus process. 
The introduced delay of the additional phase is considered insignificant compared to the time saved in view change due to the responsiveness property, as claimed by authors in~\cite{hotstuffPODC}. 

\bheading{State space.} \yu{For CHS, we instantiate the generic state $s=(cS,l_a,l_h,L)$ as follows.}
\begin{packeditemize}
    \item \textbf{Commitment state (\cs)}. The \cs represents the consecutive block structure that has already been formed. It is crucial because it directly affects the commitment of blocks. It has five possible values: $\{0, 1, 2, 3, 3'\}$, {where the integers $0$ to $3$ specifically denote the number of consecutive certified blocks.} This is because \CHS specifies that when three consecutive blocks are formed, the next block triggers a commitment. Therefore, \cs is at most 3. The special value $3'$ is used when three consecutive blocks are not continuous with the next block. In this case, the generation of the next block still triggers a commitment, but the value of \cs will reset and become 1. 

    \item \textbf{Uncommitted honest block (\lh).}  
    The notation \lh represents the number of honest blocks that have not been committed. It has three possible values: $\{0, 1, 2\}$, {tracking the count of recent blocks that the adversary can still manipulate.} \CHS locks the grandparent of the latest block, in the sense that the grandparent block and its prefix can be committed. Therefore, at most two blocks can be manipulated. {Moreover, we consider only simultaneous forks of two uncommitted honest blocks. Forking a single honest block is omitted, since the deterministic consensus rule allows the adversary to exclude both blocks from the main chain. Therefore, \lh can be safely simplified without affecting the results, and the same holds for the cases discussed below.} 
    \item \textbf{Uncommitted adversarial block (\la).} 
    The notation \la represents the number of adversarial blocks that have not been committed.
    It has two possible values: $\{0, 1\}$. The adversary can decide when honest nodes will see a block by releasing or temporarily hiding its block. Since honest nodes follow the consensus rule, once the adversary releases a certified adversarial block, it will not be overridden and can be considered as ultimately committed.

    \item \textbf{Current leader (\leader).} \yu{The leader type $L\in\{H,A\}$ follows the generic definition.}
\end{packeditemize}

\bheading{Actions.} \yu{CHS uses the generic action space
$A=\{\textsf{Adopt},\textsf{Wait},\textsf{Release},\textsf{Silent}\}$.
The availability and effect of each action are determined by the current
state and are summarized in Table~\ref{tab:state_trans_CHS}.}

\begin{table}[t]
\caption{State transition and reward matrices for \CHS. The reward of \Bh and \C is analyzed in \ssecref{subsec:CBFTmodel}.} 
\label{tab:state_trans_CHS}
\scriptsize
\centering
\ra{1.05}
\setcounter{rowID}{0} 
\scalebox{0.9}{
\begin{tabular}{@{}c@{\hspace{0.5em}}cccc@{}}
\toprule[1pt]
\textbf{Row ID} & \textbf{State $\times$ Action} & \textbf{Resulting State} & \textbf{Pr.} & \textbf{T} \\ \midrule[1pt]

\multirow{2}{*}{\nextID} & \multirow{2}{*}{\shortstack{$(\cs, 0, \lh, H)$ \\ \texttt{Adopt}}} & $(\min(\cs+1,3), 0, 1, A)$ & $\alpha$ & $\delta\!+\!2\Delta$ \\
 & & $(\min(\cs+1,3), 0, 1, H)$ & $1\!-\!\alpha$ & $3\delta$ \\ \midrule[1pt]
 
\multirow{2}{*}{\nextID} & \multirow{2}{*}{\shortstack{$(\cs, 1, \lh, H)$ \\ \texttt{Adopt}}} & $(1, 0, 1, A)$ & $\alpha$ & $\delta\!+\!2\Delta$ \\
 & & $(1, 0, 1, H)$ & $1\!-\!\alpha$ & $3\delta$ \\ \midrule[1pt]

\multirow{2}{*}{\nextID} & \multirow{2}{*}{\shortstack{$(\cs, 0, \lh, A)$ \\ \texttt{Adopt}}} & $(\cs, 1, 0, A)$ & $\alpha$ & $3\Delta$ \\
 & & $(\cs, 1, 0, H)$ & $1\!-\!\alpha$ & $\delta\!+\!2\Delta$ \\ \midrule[1pt]
 
\multirow{2}{*}{\nextID} & \multirow{2}{*}{\shortstack{$(\cs, 1, \lh, A)$ \\ \texttt{Adopt}}} & $(0/3', 1, 0, A)$ & $\alpha$ & $3\Delta$ \\
 & & $(0/3', 1, 0, H)$ & $1\!-\!\alpha$ & $\delta\!+\!2\Delta$ \\ \midrule[1pt]

\multirow{2}{*}{\nextID} & \multirow{2}{*}{\shortstack{$(\cs, 0, \lh, H)$ \\ \texttt{Wait, Silent}}} & $(\min(\cs+1,3), 0, \min(\lh+1,2), A)$ & $\alpha$ & $\delta\!+\!2\Delta$ \\
 & & $(\min(\cs+1,3), 0, \min(\lh+1,2), H)$ & $1\!-\!\alpha$ & $3\delta$ \\ \midrule[1pt]
 
\multirow{2}{*}{\nextID} & \multirow{2}{*}{\shortstack{$(\cs, 1, \lh, H)$ \\ \texttt{Wait, Silent}}} & $(1, 0, \min(\lh+1,2), A)$ & $\alpha$ & $\delta\!+\!2\Delta$ \\
 & & $(1, 0, \min(\lh+1,2), H)$ & $1\!-\!\alpha$ & $3\delta$ \\ \midrule[1pt]

\multirow{2}{*}{\nextID} & \multirow{2}{*}{\shortstack{$(\cs, 0, \lh, A)$ \\ \texttt{Wait}}} & $(0/3', 1, \lh, A)$ & $\alpha$ & $3\Delta$ \\
 & & $(0/3', 1, \lh, H)$ & $1\!-\!\alpha$ & $\delta\!+\!2\Delta$ \\ \midrule[1pt]
 
\multirow{2}{*}{\nextID} & \multirow{2}{*}{\shortstack{$(\cs, 1, 0, A)$ \\ \texttt{Wait}}} & $(\min(\cs+1,3), 1, 0, A)$ & $\alpha$ & $3\Delta$ \\
 & & $(\min(\cs+1,3), 1, 0, H)$ & $1\!-\!\alpha$ & $\delta\!+\!2\Delta$ \\ \midrule[1pt]

\multirow{2}{*}{\nextID} & \multirow{2}{*}{\shortstack{$(\cs, 1, \lh, A)$ \\ $\lh>\!0$, \texttt{Wait}}} & $(1, 1, 0, A)$ & $\alpha$ & $3\Delta$ \\
 & & $(1, 1, 0, H)$ & $1\!-\!\alpha$ & $\delta\!+\!2\Delta$ \\ \midrule[1pt]

\multirow{2}{*}{\nextID} & \multirow{2}{*}{\shortstack{$(\cs, 1, 0, H)$ \\ \texttt{Release}}} & $(\min(\cs+2,3), 0, 1, A)$ & $\alpha$ & $\delta\!+\!2\Delta$ \\
 & & $(\min(\cs+2,3), 0, 1, H)$ & $1\!-\!\alpha$ & $3\delta$ \\ \midrule[1pt]
 
\multirow{2}{*}{\nextID} & \multirow{2}{*}{\shortstack{$(\cs, 1, \lh, H)$ \\ $\lh>\!0$, \texttt{Release}}} & $(2, 0, 1, A)$ & $\alpha$ & $\delta\!+\!2\Delta$ \\
 & & $(2, 0, 1, H)$ & $1\!-\!\alpha$ & $3\delta$ \\ \midrule[1pt]

\multirow{2}{*}{\nextID} & \multirow{2}{*}{\shortstack{$(\cs, 1, 0, A)$ \\ \texttt{Release}}} & $(\min(\cs+1,3), 1, 0, A)$ & $\alpha$ & $3\Delta$ \\
 & & $(\min(\cs+1,3), 1, 0, H)$ & $1\!-\!\alpha$ & $\delta\!+\!2\Delta$ \\ \midrule[1pt]
 
\multirow{2}{*}{\nextID} & \multirow{2}{*}{\shortstack{$(\cs, 1, \lh, A)$ \\ $\lh>\!0$, \texttt{Release}}} & $(1, 1, 0, A)$ & $\alpha$ & $3\Delta$ \\
 & & $(1, 1, 0, H)$ & $1\!-\!\alpha$ & $\delta\!+\!2\Delta$ \\ \midrule[1pt]

\multirow{2}{*}{\nextID} & \multirow{2}{*}{\shortstack{$(\cs, 0, \lh, A)$\textsuperscript{a} \\ \texttt{Silent}}} & $(0, 0, \lh-1, A)$ & $\alpha$ & $2\Delta$ \\
 & & $(0, 0, \lh-1, H)$ & $1\!-\!\alpha$ & $\delta\!+\!\Delta$ \\ \midrule[1pt]

\multirow{2}{*}{\nextID} & \multirow{2}{*}{\shortstack{$(\cs, \la, \lh, A)$\textsuperscript{b} \\ \texttt{Silent}}} & $(0, 0, \lh, A)$ & $\alpha$ & $2\Delta$ \\
 & & $(0, 0, \lh, H)$ & $1\!-\!\alpha$ & $\delta\!+\!\Delta$ \\ \midrule[1pt]

\end{tabular}
}
    \begin{tablenotes}
        \item \textsuperscript{a} $\lh>\!0 \: \land \: \cs\, \neq 0/3'$
        \item \textsuperscript{b} $\la=1 \lor (\la=0 \land(\lh=0 \lor (\lh>\!0 \: \land \: \cs\, = 0/3')))$
    \end{tablenotes}
\end{table}

\bheading{State transition.} \tabref{tab:state_trans_CHS} presents a comprehensive view of all possible state transitions under each possible state, and derives each state transition, including its probability, reward and the resulting state. 

\begin{packeditemize}
    \item \textbf{Commitment state (\cs).} When the leader is honest, the transition of \cs mainly depends on \la. If $\la=\!0$, it means that there is no hidden adversarial block, and the actions \texttt{Adopt}, \texttt{Wait} and \texttt{Silent} taken by the adversary increase \cs by $1$ {(\cf Table \ref{tab:state_trans_CHS} R1 and R5)}. If $\la>\!0$, these actions signify that the adversary abandons its block. This breaks the consecutive structure and \cs resets to 0 . However, with the honest leader proposing a new block, \cs increases to $1$ {(\cf Table \ref{tab:state_trans_CHS} R2, R6)}. 
    When the adversary chooses \texttt{Release} with $\lh>\!0$, it means the released block is a forking block. This again breaks the consecutive structure, and the released block together with the new proposed honest block make \cs to $2$ {(\cf Table \ref{tab:state_trans_CHS} R11)}. If $\lh=\!0$, \cs increases by $2$ {(\cf Table \ref{tab:state_trans_CHS} R10)}.

    When the leader is adversarial, the transition can be categorized based on the action taken by the adversary. If an \texttt{Adopt} action is taken and $\la=\!0$, the adversary will extend the previous honest block, and the consecutive structure will be preserved. Also, as the proposed adversarial block is temporarily hidden, \cs remains unchanged {(\cf Table \ref{tab:state_trans_CHS} R3)}. When $\la>\!0$, the adversary takes \texttt{Adopt} action to give up the hidden block, and the consecutive structure is destroyed, and thus \cs is reset. If the previous \cs equals $3$, then \cs becomes $3'$, otherwise \cs becomes $0$ {(\cf Table \ref{tab:state_trans_CHS} R4)}. For \texttt{Wait} and \texttt{Release} actions, if a fork is formed, \cs will be reset {(\cf Table \ref{tab:state_trans_CHS} R7, R9, R11, R13)}; If the previous adversarial block is extended, \cs increases by $1$ {(\cf Table \ref{tab:state_trans_CHS} R8, R12)}. 
    If the \texttt{Silent} action is taken, \cs becomes $0$ {(\cf Table \ref{tab:state_trans_CHS} R14, R15)}.

    \item \textbf{Uncommitted adversarial block (\la).} If the current leader is adversarial and takes \texttt{Adopt}, \texttt{Wait}, or \texttt{Release} actions, it will propose a temporarily hidden adversarial block, and \la becomes $1$. In all other cases, \la becomes 0.

    \item \textbf{Uncommitted honest block (\lh).} The adversary's actions of \texttt{Adopt} and \texttt{Release} will cause \lh to become $0$, and \lh will increase by $1$ if the current leader is honest, otherwise it will remain unchanged. In the \texttt{Wait} action, if the adversary leader extends its block, the hidden block will be passively released, and \lh will be reset. Otherwise, \lh will remain unchanged. If the current leader is honest, \lh will increase by $1$. In \texttt{Silent} action, if the leader is honest, \lh will increase by $1$. If the leader is adversarial and an honest block is generated in the previous view, the block will be excluded and \lh will be reduced by $1$.


\end{packeditemize}

\bheading{Reward allocation. } \yu{We instantiate the generic reward $R(s,a,s')=(B_h,C,T)$ for CHS as follows.}


\begin{packeditemize}
    \item \textbf{Committed honest block (\Bh).} \Bh will increase by $1$ if an honest block is eventually committed on the chain. After the adversary takes \texttt{Adopt} action, the existing honest blocks will be accepted by the adversary, preventing them from being overridden, and \Bh will increase by \lh. In addition, \Bh will increase by $1$ when \lh transits to $\lh+1$ if $\lh=\!2$. Due to the locking rule, if $\lh=\!2$, the subsequent block will cause its grandparent block to be locked, and \Bh will also increase by $1$.
    
    \item \textbf{Commitment times (\C).} \C increases by $1$ for each block commitment. In \CHS, when a new block extends three consecutive blocks, it triggers a commitment. This requires \cs to be $3$ or $3'$, and meets one of the following three conditions: 1) the current leader is honest and the adversary takes \texttt{Adopt}, \texttt{Wait}, or \texttt{Silent} actions; 2) The current leader is honest, and the adversary takes \texttt{Release} action without any forks ($\lh=\!0$); or 3) The current leader is adversarial and takes \texttt{Wait} or \texttt{Release} actions without a fork ($\lh=0$).
    
    \item \textbf{Elapsed time (\T).} Recall that we use $\delta$ and $\Delta$ to denote the actual network delay and upper bound network delay after GST, respectively. The time consumed within each view consists of three parts: 1) the leader broadcasts a block, 2) the next leader collects votes from other nodes, and 3) the view-change process. 
    
    For the first two parts, they take $\delta$ if the leader is honest, and $\Delta$ otherwise. 
    For the last one, it is determined by the next leader and the responsiveness/non-responsiveness properties of the protocol. Due to the responsiveness of \CHS, the view-change process takes $\delta$ if the leader is honest and $\Delta$ otherwise.
    The leader of two consecutive views can be divided into four groups, and their consumption time differs. Note that if an adversarial leader takes the \texttt{Silent} action, honest nodes will trigger timeout upon the leader proposing the block (part 1) and directly entering the view change (part 3) without collecting votes (part 2).
\end{packeditemize}

\begin{table}[t]
\caption{State transition and reward matrices for \TCHS. The reward of \Bh and \C is analyzed in \ssecref{subsec:CBFTmodel}.} 
\label{tab:state_trans_TCHS}
\scriptsize
\centering
\ra{1.05}
\setcounter{rowID}{0}
\scalebox{0.9}{
\begin{tabular}{@{}c@{\hspace{0.5em}}ccccc@{}}
\toprule[1pt]
\textbf{Row ID} &
\textbf{State $\times$ Action} & \textbf{Resulting State} & \textbf{Pr.} & \textbf{T} \\ \midrule[1pt]
\multirow{2}{*}{\nextID} & \multirow{2}{*}{\shortstack{$(\cs, 0, \lh, H)$ \\ \texttt{Adopt}}} & $(\min(\cs+1,2), 0, 1, A)$ & $\alpha$ & $\delta\!+\!2\Delta$ \\
 & & $(\min(\cs+1,2), 0, 1, H)$ & $1\!-\!\alpha$ & $2\delta\!+\!\Delta$ \\ \midrule[1pt]
 
\multirow{2}{*}{\nextID} &\multirow{2}{*}{\shortstack{$(\cs, 1, \lh, H)$ \\ \texttt{Adopt}}} & $(1, 0, 1, A)$ & $\alpha$ & $\delta\!+\!2\Delta$ \\
& & $(1, 0, 1, H)$ & $1\!-\!\alpha$ & $2\delta\!+\!\Delta$ \\ \midrule[1pt]
\multirow{2}{*}{\nextID} &\multirow{2}{*}{\shortstack{$(\cs, 0, \lh, A)$ \\ \texttt{Adopt}}} & $(\cs, 1, 0, A)$ & $\alpha$ & $3\Delta$ \\
& & $(\cs, 1, 0, H)$ & $1\!-\!\alpha$ & $3\Delta$ \\ \midrule[1pt]
 
\multirow{2}{*}{\nextID} &\multirow{2}{*}{\shortstack{$(\cs, 1, \lh, A)$ \\ \texttt{Adopt}}} & $(0/2', 1, 0, A)$ & $\alpha$ & $3\Delta$ \\
& & $(0/2', 1, 0, H)$ & $1\!-\!\alpha$ & $3\Delta$ \\ \midrule[1pt]
\multirow{2}{*}{\nextID} &\multirow{2}{*}{\shortstack{$(\cs, 0, \lh, H)$ \\ \texttt{Wait, Silent}}} & $(\min(\cs+1,2), 0, \min(\lh+1,1), A)$ & $\alpha$ & $\delta\!+\!2\Delta$ \\
& & $(\min(\cs+1,2), 0, \min(\lh+1,1), H)$ & $1\!-\!\alpha$ & $2\delta\!+\!\Delta$ \\ \midrule[1pt]
 
\multirow{2}{*}{\nextID} &\multirow{2}{*}{\shortstack{$(\cs, 1, \lh, H)$ \\ \texttt{Wait, Silent}}} & $(1, 0, \min(\lh+1,1), A)$ & $\alpha$ & $\delta\!+\!2\Delta$ \\
& & $(1, 0, \min(\lh+1,1), H)$ & $1\!-\!\alpha$ & $2\delta\!+\!\Delta$ \\ \midrule[1pt]
\multirow{2}{*}{\nextID} &\multirow{2}{*}{\shortstack{$(\cs, 0, \lh, A)$ \\ \texttt{Wait}}} & $(0/2', 1, \lh, A)$ & $\alpha$ & $3\Delta$ \\
& & $(0/2', 1, \lh, H)$ & $1\!-\!\alpha$ & $3\Delta$ \\ \midrule[1pt]
 
\multirow{2}{*}{\nextID} &\multirow{2}{*}{\shortstack{$(\cs, 1, 0, A)$ \\ \texttt{Wait}}} & $(\min(\cs+1,2), 1, 0, A)$ & $\alpha$ & $3\Delta$ \\
& & $(\min(\cs+1,2), 1, 0, H)$ & $1\!-\!\alpha$ & $3\Delta$ \\ \midrule[1pt]

\multirow{2}{*}{\nextID} &\multirow{2}{*}{\shortstack{$(\cs, 1, \lh, A)$ \\ $\lh>\!0$, \texttt{Wait}}} & $(1, 1, 0, A)$ & $\alpha$ & $3\Delta$ \\
& & $(1, 1, 0, H)$ & $1\!-\!\alpha$ & $3\Delta$ \\ \midrule[1pt]
\multirow{2}{*}{\nextID} &\multirow{2}{*}{\shortstack{$(\cs, 1, \lh, H)$ \\ \texttt{Release}}} & $(2, 0, 1, A)$ & $\alpha$ & $\delta\!+\!2\Delta$ \\
& & $(2, 0, 1, H)$ & $1\!-\!\alpha$ & $2\delta\!+\!\Delta$ \\ \midrule[1pt]
\multirow{2}{*}{\nextID} &\multirow{2}{*}{\shortstack{$(\cs, 1, 0, A)$ \\ \texttt{Release}}} & $(\min(\cs+1,2), 1, 0, A)$ & $\alpha$ & $3\Delta$ \\
& & $(\min(\cs+1,2), 1, 0, H)$ & $1\!-\!\alpha$ & $3\Delta$ \\ \midrule[1pt]
 
\multirow{2}{*}{\nextID} &\multirow{2}{*}{\shortstack{$(\cs, 1, \lh, A)$ \\ $\lh>\!0$, \texttt{Release}}} & $(1, 1, 0, A)$ & $\alpha$ & $3\Delta$ \\
& & $(1, 1, 0, H)$ & $1\!-\!\alpha$ & $3\Delta$ \\ \midrule[1pt]
\multirow{2}{*}{\nextID} &\multirow{2}{*}{\shortstack{$(\cs, 0, \lh, A)$\textsuperscript{a} \\ \texttt{Silent}}} & $(0, 0, \lh-1, A)$ & $\alpha$ & $2\Delta$ \\
& & $(0, 0, \lh-1, H)$ & $1\!-\!\alpha$ & $2\Delta$ \\ \midrule[1pt]

\multirow{2}{*}{\nextID} &\multirow{2}{*}{\shortstack{$(\cs, \la, \lh, A)$\textsuperscript{b} \\ \texttt{Silent}}} & $(0, 0, \lh, A)$ & $\alpha$ & $2\Delta$ \\
& & $(0, 0, \lh, H)$ & $1\!-\!\alpha$ & $2\Delta$ \\ \midrule[1pt]
\end{tabular}
}
    \begin{tablenotes}
        \item \textsuperscript{a} $\lh>\!0 \: \land \: \cs\, \neq 0/2'$
        \item \textsuperscript{b} $\la=1 \lor (\la=0 \land(\lh=0 \lor (\lh>\!0 \: \land \: \cs\, = 0/2')))$
    \end{tablenotes}
\end{table}

\subsubsection{\TCHS} \TCHS shares a similar design as \CHS and Casper~\cite{casper}, and can also be viewed as the pipelined version of Tendermint~\cite{Buchman2016TendermintBF}. 
As mentioned previously, \TCHS has one less phase than \CHS, but does not have responsiveness.  
Besides, the commitment of \TCHS slightly differs from \CHS. {In \TCHS, the first block of two consecutive blocks and its previous blocks will be committed if a new block extends the second block.} In other words, in \TCHS, the parent block {of a certified block}, instead of the grandparent block, is locked.

We enumerate the state transition and reward matrices for \TCHS in \tabref{tab:state_trans_TCHS}. \yu{2CHS follows the same generic state form $s=(cS,l_a,l_h,L)$.} Due to the two consecutive blocks structure, \lh is at most 1, and \cs can be $\{0,1,2,2'\}$. 
This also leads to corresponding changes in the state transition and reward allocation of \Bh and \C. When $\lh$ increases from $1$ to $\lh+1$, \Bh will increase by $1$ {(\cf Table \ref{tab:state_trans_TCHS} R5, R6)}. When $\cs$ becomes $2$ or $2'$, honest nodes trigger a commitment, and $C$ increases by $1$ {(\cf Table \ref{tab:state_trans_TCHS} R1, R5, R8, R11)}. For time consumption, the rewards for \T in the leader-based stage remain unchanged, but since \TCHS is not responsive, the view change part always takes $\Delta$ time regardless of the identity of the next leader.
\subsubsection{\FHS} \FHS is a responsive two-phase chained BFT protocol, which allows a leader to included $2f+1$ latest QCs in its block. Compared with \CHS, the QC inclusion in \FHS slightly increases the message exchange overhead, but does not require an additional phase to achieve responsiveness. 
Despite \FHS's resilience against forking attacks, honest nodes may still lose blocks. This is because votes sent to a subsequent adversarial leader could be hidden, preventing the formation of a QC. As a result, the actions available to the adversary in \FHS are the same as \TCHS.

\begin{table}[t]
\caption{State transition and reward matrices for \FHS. The reward of \Bh and \C is analyzed in \ssecref{subsec:CBFTmodel}.} 
\label{tab:state_trans_FHS}
\scriptsize
\centering
\ra{1.05}
\setcounter{rowID}{0}
\scalebox{0.9}{
\begin{tabular}{@{}c@{\hspace{0.5em}}ccccc@{}}
\toprule[1pt]
\textbf{Row ID} & \textbf{State $\times$ Action} & \textbf{Resulting State} & \textbf{Pr.} & \textbf{T} \\ \midrule[1pt]
\multirow{2}{*}{\nextID} &\multirow{2}{*}{\shortstack{$(\cs, 0, \lh, H)$ \\ \texttt{Adopt}}} & $(\min(\cs+1,2), 0, 1, A)$ & $\alpha$ & $\delta\!+\!2\Delta$ \\
& & $(\min(\cs+1,2), 0, 1, H)$ & $1\!-\!\alpha$ & $2\delta$ \\ \midrule[1pt]
 
\multirow{2}{*}{\nextID} &\multirow{2}{*}{\shortstack{$(\cs, 1, \lh, H)$ \\ \texttt{Adopt}}} & $(1, 0, 1, A)$ & $\alpha$ & $\delta\!+\!2\Delta$ \\
& & $(1, 0, 1, H)$ & $1\!-\!\alpha$ & $2\delta$ \\ \midrule[1pt]
\multirow{2}{*}{\nextID} &\multirow{2}{*}{\shortstack{$(\cs, 0, \lh, A)$ \\ \texttt{Adopt}}} & $(\cs, 1, 0, A)$ & $\alpha$ & $3\Delta$ \\
& & $(\cs, 1, 0, H)$ & $1\!-\!\alpha$ & $2\Delta$ \\ \midrule[1pt]
 
\multirow{2}{*}{\nextID} &\multirow{2}{*}{\shortstack{$(\cs, 1, \lh, A)$ \\ \texttt{Adopt}}} & $(0/2', 1, 0, A)$ & $\alpha$ & $3\Delta$ \\
& & $(0/2', 1, 0, H)$ & $1\!-\!\alpha$ & $2\Delta$ \\ \midrule[1pt]
\multirow{2}{*}{\nextID} &\multirow{2}{*}{\shortstack{$(\cs, 0, \lh, H)$ \\ \texttt{Wait, Silent}}} & $(\min(\cs+1,2), 0, \min(\lh+1,1), A)$ & $\alpha$ & $\delta\!+\!2\Delta$ \\
& & $(\min(\cs+1,2), 0, \min(\lh+1,1), H)$ & $1\!-\!\alpha$ & $2\delta$ \\ \midrule[1pt]
 
\multirow{2}{*}{\nextID} &\multirow{2}{*}{\shortstack{$(\cs, 1, \lh, H)$ \\ \texttt{Wait, Silent}}} & $(1, 0, \min(\lh+1,1), A)$ & $\alpha$ & $\delta\!+\!2\Delta$ \\
& & $(1, 0, \min(\lh+1,1), H)$ & $1\!-\!\alpha$ & $2\delta$ \\ \midrule[1pt]
\multirow{2}{*}{\nextID} &\multirow{2}{*}{\shortstack{$(\cs, 0, \lh, A)$ \\ \texttt{Wait}}} & $(0/2', 1, \lh, A)$ & $\alpha$ & $3\Delta$ \\
& & $(0/2', 1, \lh, H)$ & $1\!-\!\alpha$ & $2\Delta$ \\ \midrule[1pt]
 
\multirow{2}{*}{\nextID} &\multirow{2}{*}{\shortstack{$(\cs, 1, 0, A)$ \\ \texttt{Wait}}} & $(\min(\cs+1,2), 1, 0, A)$ & $\alpha$ & $3\Delta$ \\
& & $(\min(\cs+1,2), 1, 0, H)$ & $1\!-\!\alpha$ & $2\Delta$ \\ \midrule[1pt]

\multirow{2}{*}{\nextID} &\multirow{2}{*}{\shortstack{$(\cs, 1, \lh, A)$ \\ $\lh>\!0$, \texttt{Wait}}} & $(1, 1, 0, A)$ & $\alpha$ & $3\Delta$ \\
& & $(1, 1, 0, H)$ & $1\!-\!\alpha$ & $2\Delta$ \\ \midrule[1pt]
\multirow{2}{*}{\nextID} &\multirow{2}{*}{\shortstack{$(\cs, 1, \lh, H)$ \\ \texttt{Release}}} & $(2, 0, 1, A)$ & $\alpha$ & $\delta\!+\!2\Delta$ \\
& & $(2, 0, 1, H)$ & $1\!-\!\alpha$ & $2\delta$ \\ \midrule[1pt]
\multirow{2}{*}{\nextID} &\multirow{2}{*}{\shortstack{$(\cs, 1, 0, A)$ \\ \texttt{Release}}} & $(\min(\cs+1,2), 1, 0, A)$ & $\alpha$ & $3\Delta$ \\
& & $(\min(\cs+1,2), 1, 0, H)$ & $1\!-\!\alpha$ & $2\Delta$ \\ \midrule[1pt]
 
\multirow{2}{*}{\nextID} &\multirow{2}{*}{\shortstack{$(\cs, 1, \lh, A)$ \\ $\lh>\!0$, \texttt{Release}}} & $(1, 1, 0, A)$ & $\alpha$ & $3\Delta$ \\
& & $(1, 1, 0, H)$ & $1\!-\!\alpha$ & $2\Delta$ \\ \midrule[1pt]
\multirow{2}{*}{\nextID} &\multirow{2}{*}{\shortstack{$(\cs, 0, \lh, A)$\textsuperscript{a} \\ \texttt{Silent}}} & $(0, 0, \lh-1, A)$ & $\alpha$ & $2\Delta$ \\
& & $(0, 0, \lh-1, H)$ & $1\!-\!\alpha$ & $2\Delta$ \\ \midrule[1pt]

\multirow{2}{*}{\nextID} &\multirow{2}{*}{\shortstack{$(\cs, \la, \lh, A)$\textsuperscript{b} \\ \texttt{Silent}}} & $(0, 0, \lh, A)$ & $\alpha$ & $2\Delta$ \\
& & $(0, 0, \lh, H)$ & $1\!-\!\alpha$ & $2\Delta$ \\ \midrule[1pt]
\end{tabular}
}
    \begin{tablenotes}
        \item \textsuperscript{a} $\lh>\!0 \: \land \: \cs\, \neq 0/2'$
        \item \textsuperscript{b} $\la=1 \lor (\la=0 \land(\lh=0 \lor (\lh>\!0 \: \land \: \cs\, = 0/2')))$
    \end{tablenotes}
\end{table}

The state transition and reward allocation of \FHS are the same as \TCHS, except for \T, which are shown in \tabref{tab:state_trans_FHS}. Besides responsiveness, \FHS optimizes the view change process by incorporating the ``happy path'' mechanism, where the next leader skips the view change if they successfully form the QC of the previous block. This QC is ensured to be the highest QC, so honest nodes can directly propose a new block based on this QC. This mechanism further reduces \T.

\section{Theoretical Results} \label{sec:MDP-eval}
In this section, we obtain the performance results of three chained BFT protocols based on our MDP model.
By observing these results, we aim to answer a key question: \textit{How do the responsiveness property and different responsiveness designs affect the \metricOne and \metricTwo of chained BFT protocols?}

To obtain the results, we cannot  directly apply our metrics to traditional MDP techniques, because the objective functions are non-linear, and the adversary aims to minimize them. Therefore, we first refer to the procedure developed by Sapirshtein et al.~\cite{sapirshtein2017optimal} to linearize the objective functions. Taking $G(\alpha)$ as an example, we first define a new objective function $G'(\alpha)$:
\begin{equation}
    G'(\alpha) =1-\mathbb{E} \left[ \liminf_{m \rightarrow \infty}{\frac{\sum_{i=1}^{m} {B_h}_{i}}{\sum_{i=1}^{m} {T_i}}} \right].
\end{equation}
The adversary's goal is to maximize $G'(\alpha)$. We use $\rho$ to denote the value of $G'(\alpha)$, and define a transformation function $w_{\rho}: \mathbb{N}^2 \to \mathbb{R}$ as:
\begin{equation}
    w_{\rho}(B_h, T) = (1-\rho) T - B_h,
\end{equation}
where $B_h$ and $T$ are the reward of the honest nodes and the time consuming in consensus process, respectively.
The reward matrix of the MDP model is determined by $w_{\rho}$. 
Suppose that $v_\rho^{\pi}$ represents the expected value under policy $\pi$, then
\begin{equation}
    v_\rho^{\pi} = \mathbb{E}\left[ \liminf_{m \rightarrow \infty} \frac{1}{m} \sum_{i=1}^{m} w_{\rho}\left({B_h}_{i}(\pi), {T}_{i}(\pi)\right) \right].
\end{equation}
Let $v_\rho^{*}$ denotes the expected value under the optimal policy, \ie, $v_\rho^{*} = \max_\pi \left\{ v_\rho^{\pi} \right\}$. According to~\cite{sapirshtein2017optimal}, if there exists some $\rho$ satisfies $v_\rho^{*}=0$, then the policy also maximizes $G'(\alpha)$.

Given that $v_0^{*}>0$ and $v_1^{*}<0$, and $v_\rho^{*}$ is monotonically decreasing in $\rho \in [0, 1]$, we can apply the binary search algorithm in~\cite{sapirshtein2017optimal} to find $\rho$ for $v_\rho^{*}=0$, denoted as $\bar{\rho}$. This $\bar{\rho}$ represents the maximum possible value of $G'(\alpha)$, which implies $1 - \bar{\rho}$ is the minimum possible value of $G(\alpha)$. The minimum value of $R(\alpha)$ can be derived in the same manner.

The range of $\alpha$ is from $0$ to $1/3$, not including $1/3$ itself (the system's maximum tolerable Byzantine nodes), with $\alpha$ increasing by $0.03$ at each step, and a precision of $10^{-4}$.
We assume that the maximum network delay $\Delta$ and the actual delay $\delta$ satisfy the relationship $\Delta\!=\!5\delta$. 
We obtain the theoretical optimal values and attack strategies through the framework, and then compare our results with existing attack strategies.
We provide the MDP source code~\footnote{MDP source code: \url{https://github.com/anonymousust/framework/tree/main/MDPModel/}.}.

\subsection{Theoretical Results Analysis}
We consider the \metricOne and \metricTwo of the protocols under optimal attack strategies found by the MDP model.
\figref{fig:growth_eval} and \figref{fig:frequency_eval}  provide the evaluation results of \metricOne and \metricTwo, respectively.

\begin{figure}[t]
    \centering
    \includegraphics[width=0.95\linewidth]{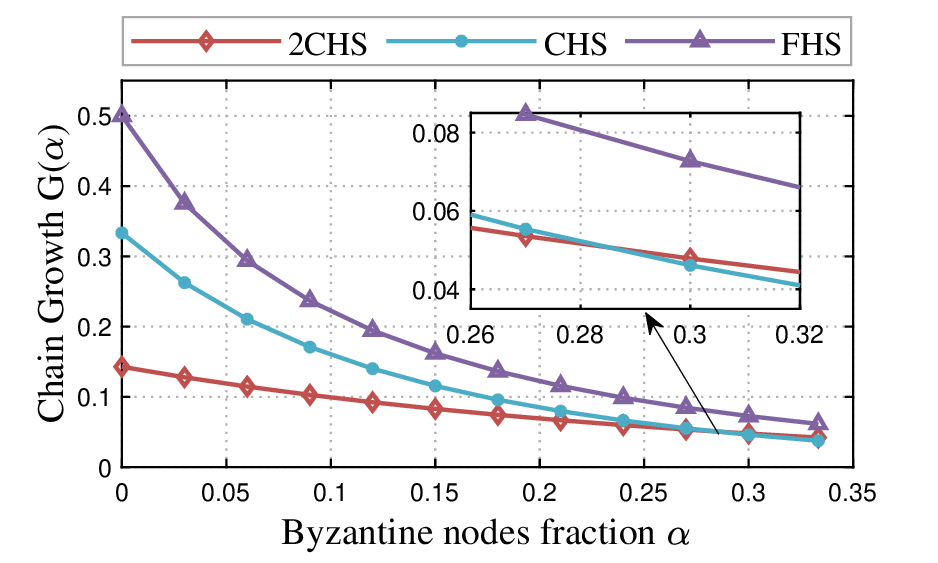}
\caption{The chain growth of \TCHS, \CHS, and \FHS. The inset zooms into the high-Byzantine-fraction region where the \TCHS and \CHS curves intersect.}
    \label{fig:growth_eval}
\end{figure}

\bheading{The performance under attacks.}
There is a downward trend in both of \metricOne and \metricTwo of the three protocols as $\alpha$ grows. \figref{fig:growth_eval} shows that, when the fraction $\alpha$ of Byzantine nodes increases from 0 to 0.3, the \metricOne of \CHS decreases from 0.333 to 0.046, and \metricOne of \FHS decreases from 0.5 to 0.073. For \TCHS, the \metricOne drops to nearly one-third when $\alpha$ is close to 0.3. 
\figref{fig:frequency_eval} shows that as $\alpha$ increases from 0 to 1/3, the \metricTwo of \CHS drops from 0.333 to 0.027, \FHS drops from 0.5 to 0.042, and \TCHS decreases from 0.143 to 0.03.  \yu{With $\delta=100$ ms, the drop of \CHS's \metricTwo from 0.333 to 0.027 translates to an increase in the average confirmation interval from approximately 300 ms to 3.7 s.} The degradation occurs because a larger $\alpha$ increases the probability of adversarial leaders, who can delay or disrupt commitment events.



\bheading{The impact of responsiveness.}
For \CHS and \FHS with responsiveness, \metricOne and \metricTwo are decreased significantly with {higher} $\alpha$. 
For example, in \FHS, when $\alpha$ increases from $0$ to $0.2$, the \metricOne decreases by $31\%$; and when $\alpha$ increases from $0.2$ to $1/3$, \metricOne decreases by $13\%$. 
In addition, with larger $\alpha$, responsiveness brings less benefits to performance: responsive \CHS and \FHS achieve comparable \metricOne and \metricTwo with non-responsive \TCHS when $\alpha$ reaches 1/3.
{When $\alpha$ is small, the impact of the adversary's attacks is limited, and the responsiveness property can effectively accelerate the view change.}
So the impact of responsiveness on performance is relatively large when $\alpha$ is small, but as $\alpha$ increases, its advantage gradually weakens.

\bheading{{Performance trade-offs across different protocol designs.}}
We observe that \FHS outperforms the other two protocols, and the non-responsive \TCHS lags behind in most cases. 
When $\alpha\!=\!0$, the \metricOne of \FHS is 3.5 and 1.5 times compared to \TCHS and \CHS, respectively. When $\alpha\!=\!1/3$, the \metricOne of \FHS becomes 1.5 and 1.6 times compared to \TCHS and \CHS. 
When $\alpha\!=\!0.3$, the \metricTwo of \FHS becomes 1.5 times compared to \TCHS and \CHS. It is noteworthy that the evaluation lines of \TCHS and \CHS intersect between $\alpha\!=\!0.27$ and $\alpha\!=\!0.3$. {Specifically, our results identify a critical efficiency threshold at this intersection: when the chain growth and commitment rate fall below approximately 0.05 and 0.04, respectively, the responsiveness of CHS no longer justifies its design complexity, as it is outperformed by the non-responsive \TCHS.} This is because of the committing rule of three consecutive blocks in \CHS.
For \metricOne, the additional phase allows the adversary to override two honest blocks in a forking attack.
For \metricTwo, triggering a block commitment is less likely compared to the rule of two consecutive blocks, especially with larger $\alpha$. {This indicates that in some cases, the protocol overhead associated with achieving responsiveness (such as adding a new phase) negates the intended performance benefits when under attack.} This is also evidenced in \FHS that achieves better \metricOne and \metricTwo due to its optimization for the happy paths. {These observations underscore that protocol performance is an interplay of responsiveness and specific mechanisms. }

\begin{figure}[t]
    \centering
    \includegraphics[width=0.95\linewidth]{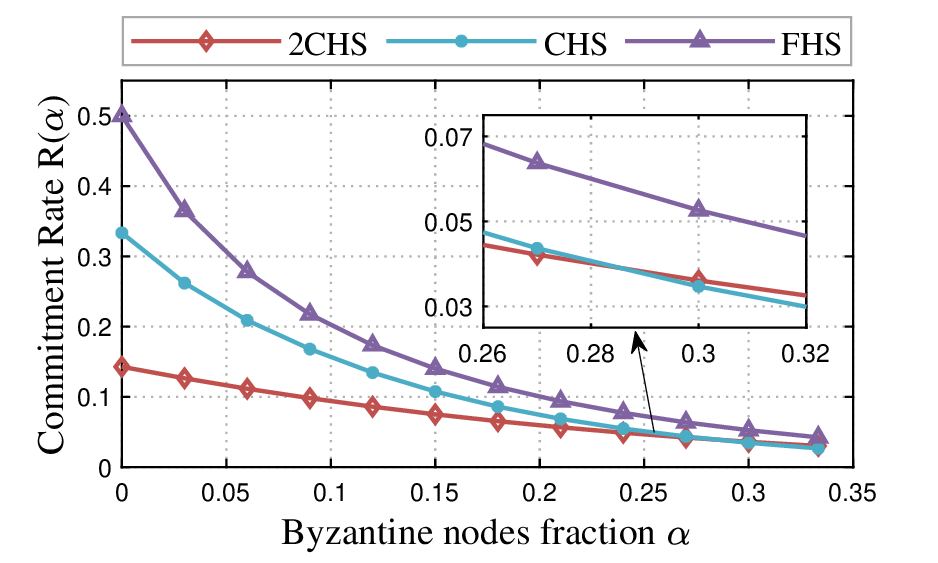}
    \caption{The commitment rate of \TCHS, \CHS, and \FHS. The inset zooms into the high-Byzantine-fraction region where the \TCHS and \CHS curves intersect.}
    \label{fig:frequency_eval}
\end{figure}

\bheading{Considerations for Protocol Design.}
\yu{The extra communication phase in \CHS and the latest QC proof in \FHS, while supporting responsive designs, also increases message overhead and processing time.} This can exacerbate transmission delays and the risk of network congestion, particularly impacting nodes with limited resources by slowing their message processing and consensus participation. In networks with a higher fraction of Byzantine nodes, these negative effects of responsiveness become more pronounced.
Therefore, when implementing responsiveness in protocol design, it is crucial to consider its benefits against potential drawbacks, especially in networks with Byzantine nodes.

\bheading{\yu{Summary and design implications.}} 
\yu{The evaluation shows that attacks can significantly degrade both chain growth and commitment rate, especially when the fraction $\alpha$ is large. Furthermore, achieving responsiveness can improve chain growth and commitment rate when $\alpha$ is small. However, when $\alpha$ is relatively large, the performance advantage of responsive designs becomes smaller. This reveals that responsiveness is not a cost-free performance enhancement, but rather a design trade-off.}

\yu{From a protocol-design perspective, the results indicate that the mechanisms introduced to achieve responsiveness should be evaluated together with their attack surface. For example, the additional phase and three-consecutive-block
committing rule in CHS may reduce its advantage under stronger attacks, whereas the happy-path optimization in FHS helps preserve better performance. Therefore, appropriate implementation needs to be considered based on protocol needs, especially in the presence of attacks.}


\subsection{Comparison to Existing Attack Strategies}

Modeling chained BFT protocols as MDPs allows us to derive optimal adversarial strategies and evaluate the corresponding worst-case performance metrics. To demonstrate the effectiveness of our approach, we compare our strategies against the baseline strategy in~\cite{giridharan2023beegees, gai2021dissecting}, as no other comparable baselines are available. Using this strategy, adversarial leaders remain silent during their views, forcing a timeout in the proposal phase and triggering a view change for all nodes. This behavior resets the consecutive block count, effectively preventing block commitments and degrading system performance.

\figref{fig:compareCHS} illustrates the commitment rate of \CHS under different attack strategies. Both the simple attack strategy and our proposed strategy result in a significant reduction in \metricTwo compared to scenarios without attacks. Notably, the impact becomes more pronounced as $\alpha$ increases. For instance, when $\alpha = 0.3$, \metricTwo under the simple attack strategy and our strategy drops to $12\%$ and $10\%$, respectively, of the commitment rate without attacks. 

\bheading{{Optimal adversarial actions.}} {We identify that the most detrimental adversarial impact arises from a strategic combination of actions introduced in Section IV. By identifying the optimal timing and targets for these actions, our strategy consistently outperforms the simple attack approach, particularly in corner cases.} For example, when an adversarial node is elected as the leader, and no three-consecutive block structure exists, the adversary can propose a block but selectively send the proposal to a subset of honest nodes. {Unlike a simple \texttt{Silent} attack where all nodes timeout simultaneously at the proposal phase, this tactic forces a subset of nodes to progress into the vote collection stage. Since these nodes reset their local timers upon receiving the proposal, they must wait through an additional timeout period when the required QC fails to form. This staggered timeout mechanism creates temporary asynchrony among honest nodes, delaying the global transition to the next view.} {We categorize this behavior as the selective release tactic.} This tactic prevents honest nodes from gathering enough votes to form a QC, {thereby maximizing the consumed time and resetting the number of consecutive blocks.} Consequently, the adversary resets the number of consecutive blocks while extending the consumed time. This sophisticated approach amplifies the attack's impact on \metricTwo by leveraging the protocol's inherent vulnerabilities. {This discovery represents more than a mere empirical observation; it identifies a formal vulnerability in the chained commit rules and establishes the theoretical lower bound of protocol responsiveness by systematically exploring the worst-case scenario via MDP.} Since the case shows with small probability, the performance gap between the baseline and our strategy remains moderate. Nevertheless, the powerful MDP framework enables us to find the optimal strategy for more complex chained BFT protocols, advancing existing attack methods.

\begin{figure}[t]
    \centering
    \includegraphics[width=0.95\linewidth]{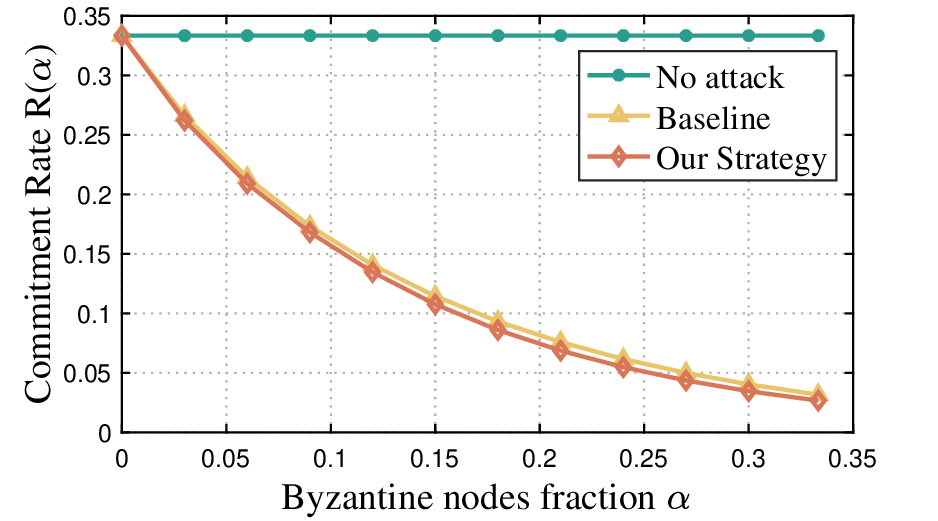}
    \caption{The \metricTwo of \CHS under our strategy, the baseline(simple strategy) and without attack.}
    \label{fig:compareCHS}
\end{figure}

\begin{figure*}[h]
    \centering
    \begin{subfigure}{0.32\textwidth}
        \centering
        \includegraphics[width=0.9\textwidth]{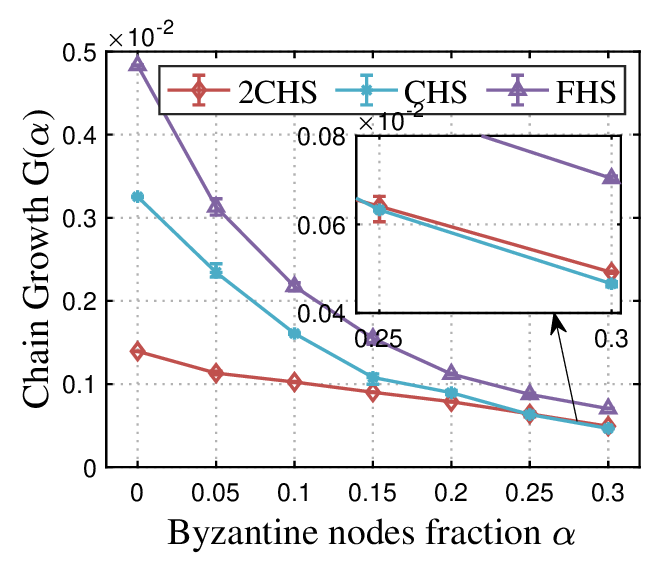}
        \caption{Chain growth when $\delta=100ms$.}
        \label{fig:growth0_experiment}
    \end{subfigure} 
    \hfill
    \begin{subfigure}{0.32\textwidth}
        \centering   
        \includegraphics[width=0.9\textwidth]{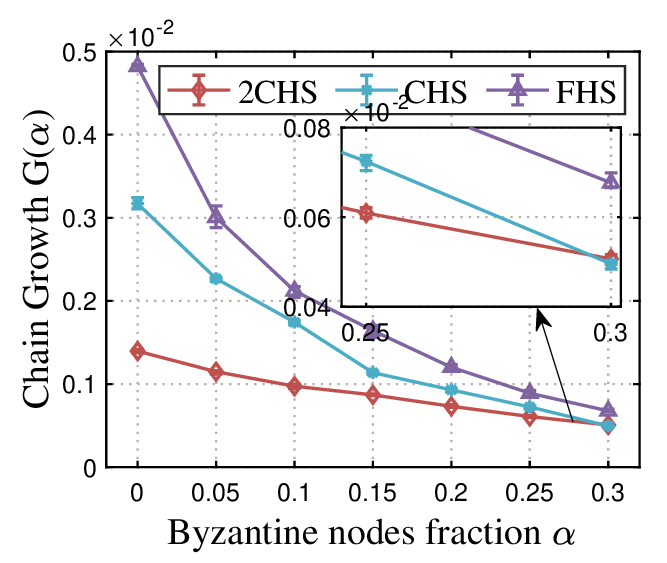}
        \caption{Chain growth when $\delta=100\pm25ms$.}
        \label{fig:growth1_experiment}
    \end{subfigure}
    \hfill
    \begin{subfigure}{0.32\textwidth}
        \centering   
        \includegraphics[width=0.9\textwidth]{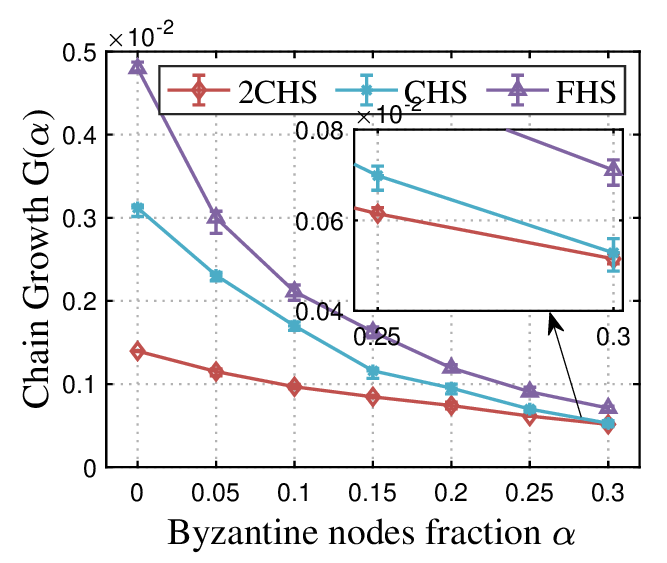}
        \caption{Chain growth when $\delta=100\pm50ms$.}
        \label{fig:growth2_experiment}
    \end{subfigure}
    \hfill
    \begin{subfigure}{0.32\textwidth}
        \centering
        \includegraphics[width=0.9\textwidth]{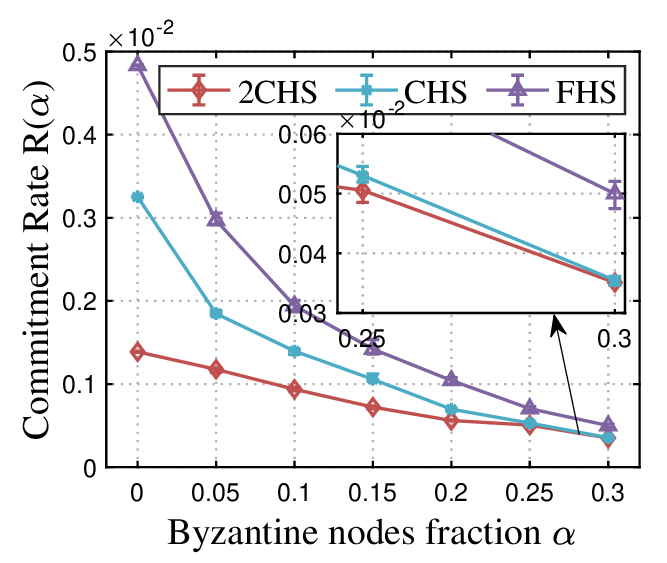}
        \caption{Commitment rate when $\delta=100ms$.}
        \label{fig:rate0_experiment}
    \end{subfigure} 
    \hfill
    \begin{subfigure}{0.32\textwidth}
        \centering   
        \includegraphics[width=0.9\textwidth]{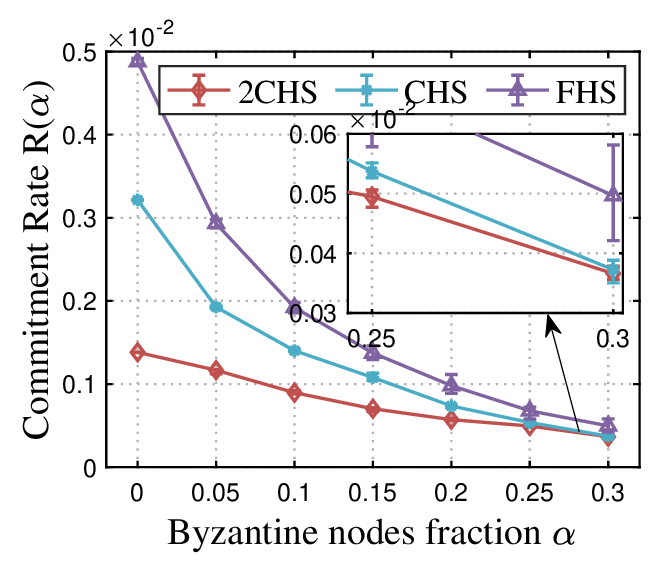}
        \caption{Commitment rate when $\delta=100\pm25ms$.}
        \label{fig:rate1_experiment}
    \end{subfigure}
    \hfill
    \begin{subfigure}{0.32\textwidth}
        \centering   
        \includegraphics[width=0.9\textwidth]{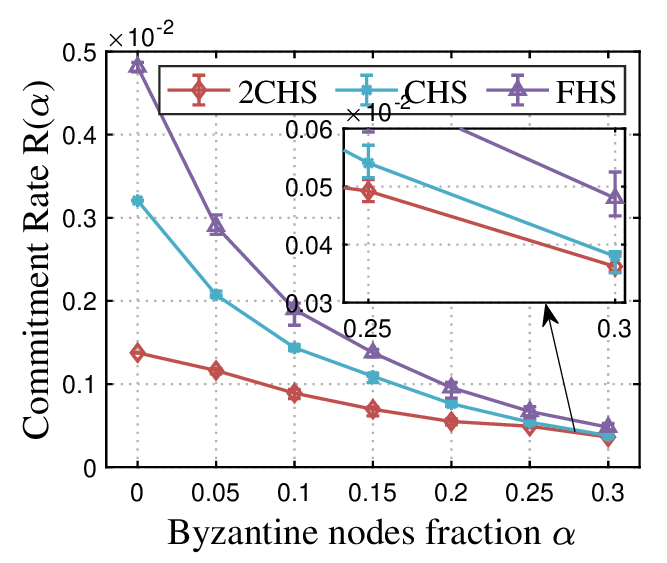}
        \caption{Commitment rate when $\delta=100\pm50ms$.}
        \label{fig:rate2_experiment}
    \end{subfigure}
    \caption{The experimental results of \metricOne and \metricTwo of \TCHS, \CHS, and \FHS with varying network delay $\delta$.}
\label{fig:experiment-growth}
\end{figure*}

\section{Experiments} \label{sec:experiments}
The theoretical models provide a foundation for understanding the impact of responsiveness on system performance under various conditions. 
To study the practical implications of the theoretical results, we implement the three studied chained BFT protocols and evaluate their \metricOne and \metricTwo in real-world scenarios. 
Besides, we consider more factors of practical systems, such as fluctuating network delay in evaluation. 
The adversary in our experiments employs the attack strategies found by MDP in \ssecref{sec:MDP-eval}. The experiment will check whether the experimental results align with the theoretical results and examine the impact of fluctuating network delay $\delta$ on the results.

\subsection{Implementation and Setup}

\bheading{Implementation.} We have extended an open-source evaluation platform Bamboo\footnote{Bamboo: \url{https://github.com/gitferry/bamboo}.} in the Go language, which already supports \CHS, \TCHS, and \FHS protocols.
Our modifications\footnote{Attack implementation based on Bamboo: \url{https://github.com/anonymousust/framework/tree/main/experiment}} add approximately 1000 lines of code to Bamboo for implementing attack strategies on \CHS, \TCHS, and \FHS.

\bheading{System setup.} {We deploy our system on AWS EC2 machine for experiments in realistic scenarios.} The experimental environment includes 4 servers, each equipped with a 16-core CPU, 32GB RAM, with the operating system of Ubuntu Server 22.04. The network latency between servers is approximately 1ms.
The experimental environment consists of a fixed number of nodes, with the flexibility to control the fraction of Byzantine nodes within the network. Our experimental settings follow conventional BFT configurations to experimentally validate the theoretical findings. Specifically, we set up a network topology that could accommodate up to 60 nodes (each server supports 15 nodes), with the capability to introduce up to 18 Byzantine nodes to test the system's performance under attacks.
The key parameters for the experiments include the fraction of Byzantine nodes and the network latency settings. {Recall that we assume a saturated workload, which implies that client transactions are continuously being processed. Therefore, we do not consider factors such as variable transaction arrival rates or partial synchrony.} 

To align with the theoretical setting, \ie, $\Delta=5\delta$, we first set the minimum delay between nodes at 100 ms and the maximum latency for communication at 500 ms. 
Meanwhile, to demonstrate the impact of a random network environment, we set the minimum delay between nodes to a fluctuating value between 25 ms and 50 ms, and conduct corresponding experiments based on the previous ones. The fluctuation of delay conforms to a uniform distribution.
We run each experiment for 10 minutes and derive the average results over six independent runs. 

\subsection{Chain Growth}

We first use a fixed network delay $\delta\!=\!100 ms$ in experiments. 
\figref{fig:growth0_experiment} shows that the experimental results closely match the results in \ssecref{sec:MDP-eval}. 
The \metricOne of \CHS, \TCHS, and \FHS gets worse as $\alpha$ increases. 
When $\alpha\!=\!0.3$, \metricOne of \TCHS drops to $35\%$ of that when $\alpha\!=\!0$. 
\FHS outperforms \TCHS and \CHS in \metricOne, and \TCHS and \CHS have an intersection point.

In the presence of network delay fluctuations, as shown in \figref{fig:growth1_experiment} and \figref{fig:growth2_experiment}, we observe that the performance of \metricOne remains largely consistent with the stable delay scenario. However, the measurement errors indicated by the error bars increase with the delay fluctuations. In the case of \CHS, with a fluctuation range of 25ms, the measurement error is 1.3 times compared to the stable delay, and with a fluctuation range of 50ms, the error doubles compared to without fluctuation. 
Overall, the experimental results of \metricOne with stable delay are consistent with the theoretical results, thus confirming the findings in \ssecref{sec:MDP-eval}.

\subsection{Commitment Rate}
\figref{fig:rate0_experiment} shows the \metricTwo results under stable network delay. The \metricTwo also shows a decline as $\alpha$ grows, with \FHS outperforming \TCHS and \CHS. As $\alpha$ increases from 0 to 0.3, \metricTwo of \FHS decreases by $11\%$. 
\figref{fig:rate1_experiment} and \figref{fig:rate2_experiment} show the evaluation results with a delay fluctuation amplitude of 25 and 50 ms. 
The \metricTwo is similar to that with stable delay.
The measurement errors increase with increasing fluctuations. Regarding \TCHS, at fluctuation ranges of 25ms and 50ms, the measurement errors are 1.2 times and 2.1 times compared to those without fluctuations, respectively. For \FHS, the measurement errors become 2 and 2.3 times of the original values, respectively. \CHS and \FHS show more significant deviations compared to \TCHS.
While the measurement error increases with more fluctuating network latency, the average evaluation results remain largely unchanged, and the observed trend is consistent with previous experiments.


\yu{In a nutshell, the experiments provide direct evidence supporting the validity of the framework. Under both stable and fluctuating-delay settings, the main trends of chain growth and commitment rate remain consistent with the theoretical results. Larger delay fluctuations mainly increase the measurement errors, while the average evaluation results remain largely unchanged. This shows that the conclusions from the framework remain robust under the considered fluctuating network conditions.}

\begin{table}[t]
\caption{State transition and reward matrices for \Streamlet.} 
\label{tab:state_trans_Streamlet}
\scriptsize
\centering
\ra{1.05}
\begin{tabular}{@{}ccccc@{}}
\toprule[1pt]
\textbf{State $\times$ Action} & \textbf{Resulting State} & \textbf{Pr.} & \textbf{\Bh} & \textbf{\C} \\ \midrule[1pt]
\multirow{2}{*}{\shortstack{$(\cs, 0, \lh, H)$ \\ \texttt{Adopt}}} & $(\min(\cs+1,3), 0, 1, A)$ & $\alpha$ & \multirow{2}{*}{\lh} & \multirow{2}{*}{\shortstack{if $\cs=2/3$\\ \C=1}} \\
 & $(\min(\cs+1,3), 0, 1, H)$ & $1\!-\!\alpha$ & & \\ \midrule[1pt]
 
\multirow{2}{*}{\shortstack{$(\cs, 1, \lh, H)$ \\ \texttt{Adopt}}} & $(1, 0, 1, A)$ & $\alpha$ & \multirow{2}{*}{\lh} & \multirow{2}{*}{0} \\
 & $(1, 0, 1, H)$ & $1\!-\!\alpha$ &  & \\ \midrule[1pt]
\multirow{2}{*}{\shortstack{$(\cs, 0, \lh, A)$ \\ \texttt{Adopt}}} & $(\cs, 1, 0, A)$ & $\alpha$ & \multirow{2}{*}{\lh} & \multirow{2}{*}{0} \\
 & $(\cs, 1, 0, H)$ & $1\!-\!\alpha$ & & \\ \midrule[1pt]
 
\multirow{2}{*}{\shortstack{$(\cs, 1, \lh, A)$ \\ \texttt{Adopt}}} & $(0, 1, 0, A)$ & $\alpha$ & \multirow{2}{*}{\lh} & \multirow{2}{*}{0} \\
 & $(0, 1, 0, H)$ & $1\!-\!\alpha$ & & \\ \midrule[1pt]
\multirow{2}{*}{\shortstack{$(\cs, 0, \lh, H)$ \\ \texttt{Wait, Silent}}} & $(\min(\cs+1,3), 0, \lh+1, A)$ & $\alpha$ & \multirow{2}{*}{0} & \multirow{2}{*}{\shortstack{if $\cs=2/3$\\ \C=1}} \\
 & $(\min(\cs+1,3), 0, \lh+1, H)$ & $1\!-\!\alpha$ & & \\ \midrule[1pt]
 
\multirow{2}{*}{\shortstack{$(\cs, 1, \lh, H)$ \\ \texttt{Wait, Silent}}} & $(1, 0, \lh+1, A)$ & $\alpha$ & \multirow{2}{*}{0} & \multirow{2}{*}{0} \\
 & $(1, 0, \lh+1, H)$ & $1\!-\!\alpha$ & & \\ \midrule[1pt]
\multirow{2}{*}{\shortstack{$(\cs, 0, \lh, A)$ \\ \texttt{Wait}}} & $(0, 0, \lh, A)$ & $\alpha$ & \multirow{2}{*}{0} & \multirow{2}{*}{0} \\
 & $(0, 0, \lh, H)$ & $1\!-\!\alpha$ & & \\ \midrule[1pt]

\multirow{2}{*}{\shortstack{$(\cs, 1, \lh, A)$ \\  \texttt{Wait}}} & $(\min(\cs+1,3), 1, 0, A)$ & $\alpha$ & \multirow{2}{*}{\lh} & \multirow{2}{*}{\shortstack{if $\cs=2/3$\\ \C=1}} \\
 & $(\min(\cs+1,3), 1, 0, H)$ & $1\!-\!\alpha$ & & \\ \midrule[1pt]
\multirow{2}{*}{\shortstack{$(\cs, 1, \lh, H)$ \\ \texttt{Release}}} & $(\min(\cs+2,3), 0, 1, A)$ & $\alpha$ & \multirow{2}{*}{\lh} & \multirow{2}{*}{(1)} \\
 & $(\min(\cs+2,3), 0, 1, H)$ & $1\!-\!\alpha$ & & \\ \midrule[1pt]

\multirow{2}{*}{\shortstack{$(\cs, 1, \lh, A)$ \\ \texttt{Release}}} & $(\min(\cs+1,3), 1, 0, A)$ & $\alpha$ & \multirow{2}{*}{\lh} & \multirow{2}{*}{\shortstack{if $\cs=2/3$\\ \C=1}} \\
& $(\min(\cs+1,3), 1, 0, H)$ & $1\!-\!\alpha$ & & \\ \midrule[1pt]

\multirow{2}{*}{\shortstack{$(\cs, 1, \lh, H)$ \\ \texttt{Withhold}}} & $(0, 0, 0, A)$ & $\alpha$ & \multirow{2}{*}{\lh} & \multirow{2}{*}{0} \\
 & $(0, 0, 0, H)$ & $1\!-\!\alpha$ & & \\ \midrule[1pt]

\multirow{2}{*}{\shortstack{$(\cs, 1, \lh, A)$ \\ \texttt{Withhold}}} & $(\min(\cs+1,3), 1, 0, A)$ & $\alpha$ & \multirow{2}{*}{\lh} & \multirow{2}{*}{\shortstack{if $\cs=2/3$\\ \C=1}} \\
& $(\min(\cs+1,3), 1, 0, H)$ & $1\!-\!\alpha$ & & \\ \midrule[1pt]

\multirow{2}{*}{\shortstack{$(\cs, \la, \lh, A)$ \\ \texttt{Silent}}} & $(0, 0, 0, A)$ & $\alpha$ & \multirow{2}{*}{0} & \multirow{2}{*}{0} \\
 & $(0, 0, 0, H)$ & $1\!-\!\alpha$ & & \\ \midrule[1pt]

\end{tabular}
    \begin{tablenotes}
        \item \text{(1)} If \cs=1, \C=1; if \cs=2/3, \C=2
    \end{tablenotes}
\end{table}

\section{Beyond Chosen Protocols} \label{sec:beyond}
We primarily select the classic protocols that have been deployed in real distributed systems and influence subsequent BFT protocols~\cite{neiheiser2021kauri}. At the same time, there are also some new chained BFT protocols, such as Streamlet~\cite{chan2020streamlet}, Marlin~\cite{sui2022marlin}, and BeeGees~\cite{giridharan2023beegees}, that differ in design (see \ssecref{subsec:cbftProtocols}) from three protocols we analyzed.
Nevertheless, our modeling framework could be applied to these protocols with slight modifications because they share the same consensus paradigm. {Furthermore, the emergence of newer protocol designs, such as  HotStuff-2~\cite{malkhi2023HotStuff} and Wendy~\cite{giridharan2021no}, has specifically targeted responsiveness without incurring an extra phase. These optimized design considerations are highly consistent with our \textit{Finding 2}, which states that the design must balance responsiveness with protocol-specific requirements by considering adversarial behaviors. Our MDP-based framework provides the quantitative tools necessary to evaluate such protocol evolutions. By establishing a rigorous baseline for classic chained protocols, our model offers a general methodology to analyze the responsiveness-overhead trade-offs in both established and emerging BFT frontiers.}

We extended our framework to Streamlet.
\Streamlet is a streamlined chained BFT protocol that is different from \TCHS, \CHS and \FHS.
Firstly, it follows the longest certified chain rule, where an honest node only votes to the block that extends the longest certified chain. Due to this rule, the forking attack described in the main text will not appear in \Streamlet. If the adversary tries to override an honest block, the honest nodes will not vote for it. However, the longest certified chain rule also introduces another way of attack. If the adversarial leader sends its proposal of block $B$ to only part of the honest nodes, excluding the next leader. The next leader does not see $B$ and thus proposes a block $B'$ at the same height of $B$. After this new proposal, the adversarial nodes release their votes on $B$. At this point, honest nodes that previously received $B$ will consider it the longest chain and will not vote for $B'$. We can consider this type of attack as a preemptive forking attack. This difference will add a new action to the action space $A$.

Secondly, the lock and commit rules of \Streamlet (referred to as notarize and final respectively in \Streamlet) are different from the previous three protocols. Due to the broadcasting vote pattern, nodes can collect votes and generate QC locally at the end of each view. 
The block will be locked once certified. If a structure of three consecutive blocks is formed, the first two blocks will be committed. This difference will affect the reward allocation for \Bh and \C. Moreover, \Streamlet does not have view change steps and is not responsive, so the consumed time in each view is the same, i.e. 2$\Delta$. This mainly affects the reward distribution of \T. The MDP modeling of \Streamlet is shown in \tabref{tab:state_trans_Streamlet} in detail.

We also use the framework to evaluate the chain growth and commitment rate of \Streamlet, and the results are shown in \figref{fig:Streamlet-eval}. Due to the lack of responsiveness, both metrics decrease relatively smoothly as $\alpha$ increases. When $\alpha$ increases from $0$ to $0.3$, the \metricOne decreases from $0.1$ to $0.049$, and \metricTwo decreases from $0.1$ to $0.028$. Compared to the other three protocols discussed earlier, the design of Streamlet is simpler, but its overall performance will also be slightly worse.

\begin{figure}[t]
    \begin{subfigure}{0.2416\textwidth}
        \centering
        \includegraphics[width=\linewidth]{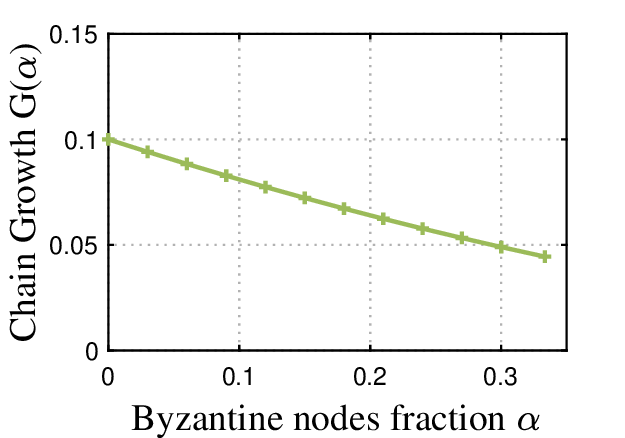}
        \caption{The \metricOne.}
        \label{fig:Streamlet-growth}
    \end{subfigure}
    \begin{subfigure}{0.2416\textwidth}
        \centering
        \includegraphics[width=\linewidth]{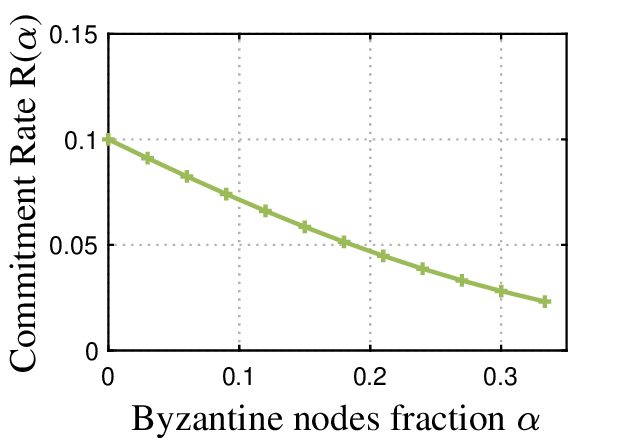}
        \caption{The \metricTwo.}
        \label{fig:Streamlet-rate}
    \end{subfigure}
    \caption{The \metricOne and \metricTwo of \Streamlet.}
    \label{fig:Streamlet-eval}
\end{figure}

\section{Discussion}\label{sec:discussion}
\yt{We discuss the robustness of our results under relaxed modeling assumptions, the limitations and future extensions of our current approach, and the implications for practical deployments and protocol design.}

\subsection{Modeling Assumptions and Robustness}
\bheading{Adversary coordination delay.} 
\yu{We simplify the adversary's behaviors in the MDP modeling to analyze the worst-case scenario. 
For example, we assume the adversary coordinates all Byzantine nodes without additional coordination delay, which may not hold in dynamic network conditions. 
Such coordination delay would reduce the adversary's feasible actions and weaken timing-sensitive attacks such as selective release. 
Thus, while the practical effect of a strategy may change under such constraints, our analytical results can still be interpreted as lower bounds on chain growth and commitment rate under the modeled strong adversary.}

\bheading{Network delay distribution.}
\yu{Real-world WANs may exhibit non-uniform delay distributions due to packet loss, routing changes, and heterogeneous links. 
Such delays may cause honest nodes to miss voting or view-change windows, while also making adversarial timing less predictable. 
They may change the exact values and variance of chain growth and commitment rate, but do not alter the protocol-level trends captured by our model.}

\bheading{Defense mechanisms.} We notice that some latest BFT protocols consider accountability, forensics, and reputation support~\cite{sheng2021bft}, in which honest nodes can detect certain attacks and identify Byzantine nodes. 
\yt{Such defenses can mitigate attacks through punishment, but they are unlikely to fully eliminate the threat, since economically rational adversaries may operate below the slashing/reputation threshold or launch attacks intermittently, and attackers indifferent to penalties are unconstrained. }

\subsection{Limitations and Future Extensions}
\bheading{Asynchronous evaluation.} Our analysis focuses on synchronous periods after GST, where message delivery between nodes is confined within a known bound. The attacks on protocol performance may be more severe in asynchronous situations where message delivery time is unpredictable, and network partitions can occur. 
Therefore, modeling and evaluating chained BFT protocols in asynchronous periods (\ie, before GST) may reveal more insights about the impact of responsiveness.

\bheading{Incentive-based attacks.} We do not analyze the impact of these attacks on nodes' rewards in this paper. This is because incentive-based attacks are tightly coupled with the underlying reward mechanism. In other words, a rigorous analysis requires a concrete reward design. However, as no practical system currently provides such a design, we do not analyze these attacks in this work and leave them for future exploration.

\bheading{Adaptive adversary.} \yu{Our current analysis assumes a fixed Byzantine fraction and does not model adaptive corruption, where the adversary may corrupt new nodes or adjust its control according to the observed leader schedule and protocol history. 
Such an adversary may further degrade protocol performance. 
Our MDP framework can be extended to capture this setting by incorporating leader-history information or time-varying adversarial control into the state space.}

\subsection{Practical and Design Implications}
\bheading{Practical implications for blockchain systems.}
\yt{In practical blockchain deployments, leader selection is typically stake-weighted rather than random. However, our MDP model remains applicable because the parameter $\alpha$ can be interpreted as the fraction of the total stake controlled by the adversary.} For protocols like Aptos that employ leader reputation systems to counter such adversaries, our identified strategies (\eg, selective release) provide a theoretical benchmark for evaluating these defenses, and subtle commitment stalls can serve as a useful signal for demoting leaders.

\bheading{Design implications for responsiveness.}
\yu{Our findings suggest that responsiveness should be considered together with protocol complexity and deployment conditions. 
When the Byzantine fraction is low and the network is stable, mechanisms for achieving responsiveness can improve performance. 
However, under stronger adversarial influence or more dynamic network conditions, these mechanisms should be evaluated together with their added protocol complexity, timeout behavior, view-change logic, and potential attack surfaces.}

\section{Conclusion} \label{sec:conclusion}
In this paper, we analyze the performance of chained BFT protocols through a unified framework with two metrics, \metricOne and \metricTwo. This framework is constructed upon an MDP model, which allows us to evaluate the worst-case performance of three representative protocols (\ie, \CHS, \TCHS, and \FHS) under attacks. {The MDP model is tailored to chained BFT consensus via state space redefinition, linearization optimization, and network delay integration, filling the gap of quantitative MDP-based evaluation for non-PoW consensus.}
The evaluation reveals the effect of responsiveness, which is an important property in protocol design. The framework also provides the optimal attack strategies, discovering better strategies than previous work, which can further decrease \metricTwo. We also conduct an experimental deployment and confirm that the experimental results of our metrics under optimal strategies on practical platforms align well with our theoretical results. In addition, the framework can be extended to other chained BFT protocols and provides a convenient tool for performance comparison among these protocols.

\normalem
\bibliographystyle{IEEEtran}
\bibliography{data/reference}

\vskip -2\baselineskip plus -1 fil
\begin{IEEEbiography}[{\includegraphics[width=1in,height=1in,clip,keepaspectratio]{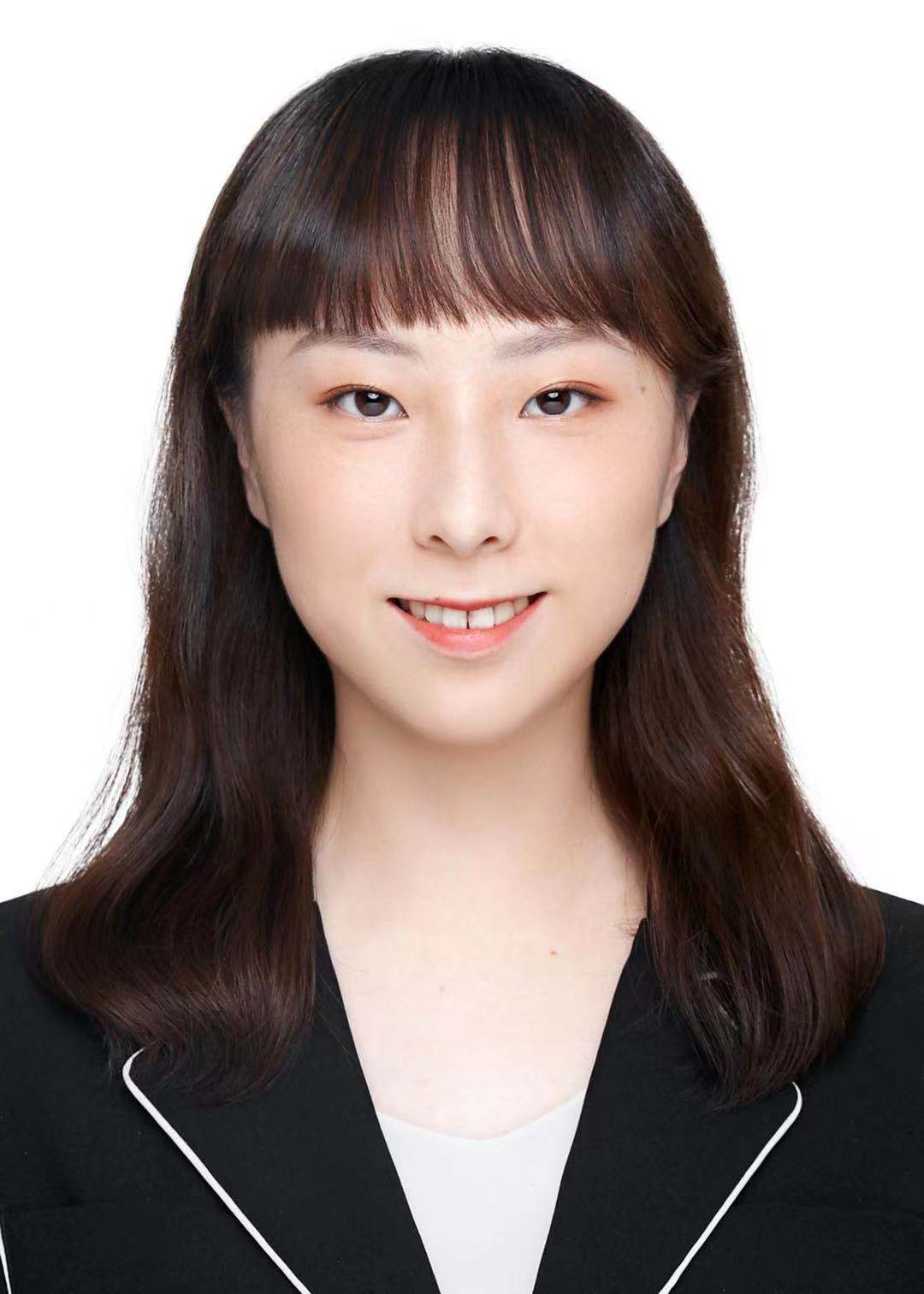}}] {Yining Tang} is currently a
Researcher with the Research Institute, China Telecom Company Ltd. She received her M.E. degree from Southern University of Science and Technology, in 2025. Her research interests include blockchain and network security.
\end{IEEEbiography}

\vskip -2\baselineskip plus -1 fil
\begin{IEEEbiography}[{\includegraphics[width=1in,height=1in,clip,keepaspectratio]{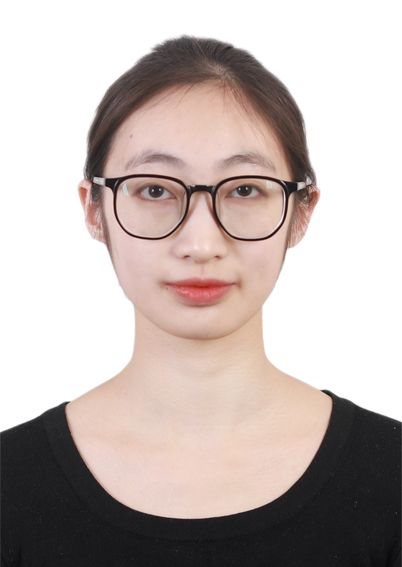}}] {Mohan Yu} is currently a visiting student at Southern University of Science and Technology. She received her M.E. degree from Southeast University. Her research interests include blockchain and network security.
\end{IEEEbiography}

\vskip -2\baselineskip plus -1 fil
\begin{IEEEbiography}[{\includegraphics[width=1in,height=1in,clip,keepaspectratio]{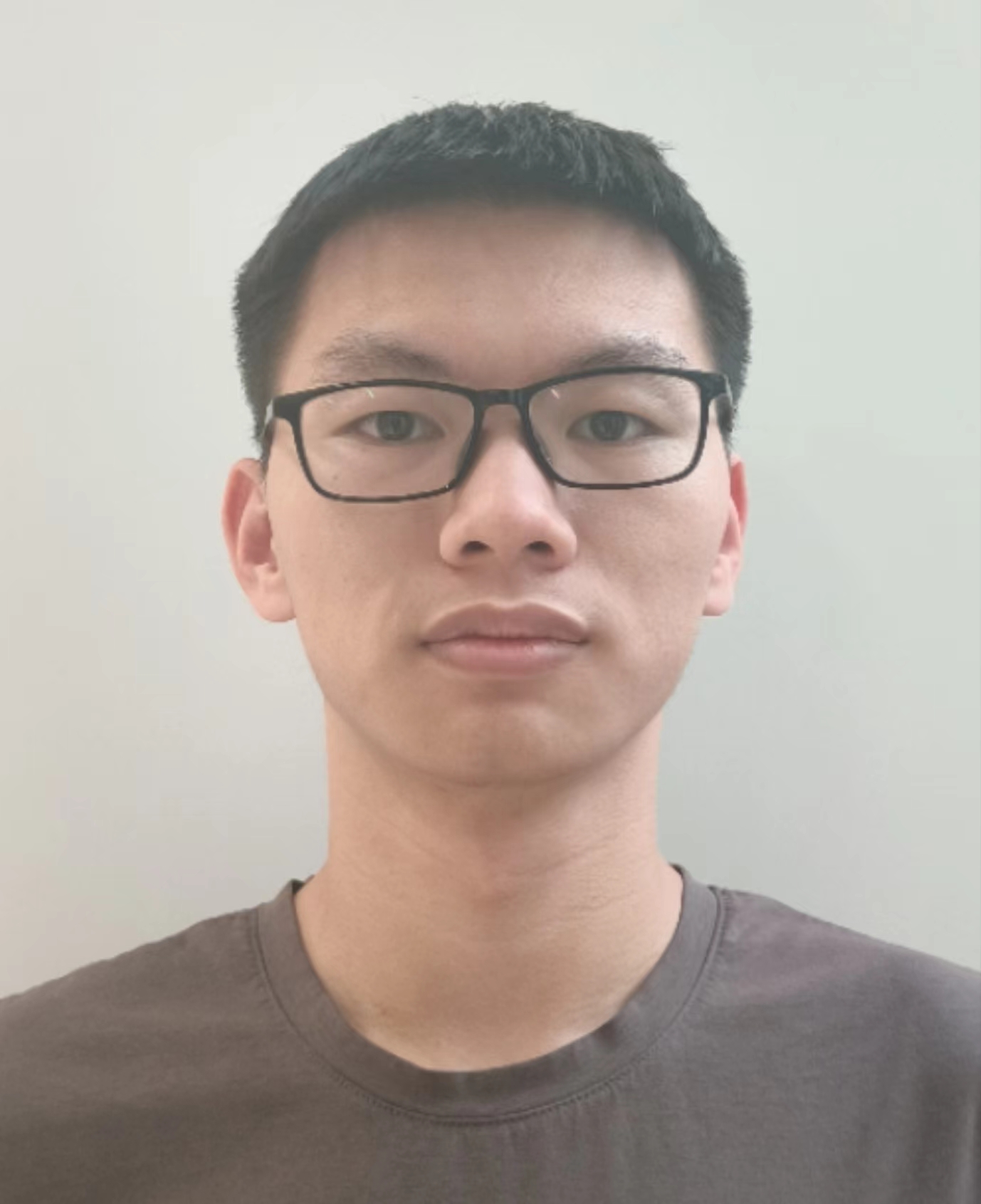}}] {Qihang Luo} is currently pursuing the B.E. degree from Southern University of Science and Technology. His research interests include blockchain and trusted execution environments.
\end{IEEEbiography}

\vskip -2\baselineskip plus -1 fil
\begin{IEEEbiography}[{\includegraphics[width=1in,height=1.15in,clip,keepaspectratio]{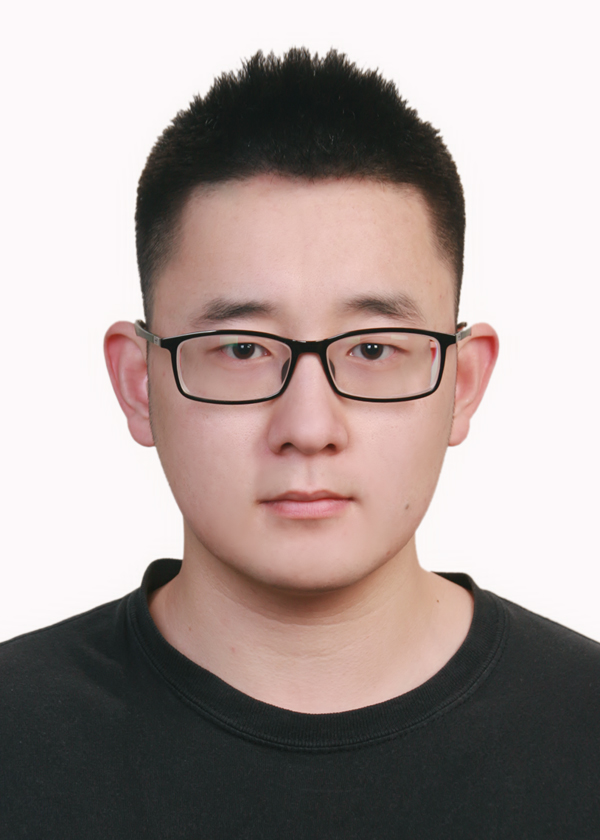}}]{Runchao Han} is currently a Researcher and Engineer at Babylon Labs. He completes his PhD at Monash University and CSIRO's Data61, Australia. He received an MSc degree from The University of Manchester, United Kingdom, and a bachelor degree from Beijing University of Posts and Telecommunications, China.
His research focuses on distributed systems, especially security and scalability issues in blockchains. 
\end{IEEEbiography}

\vskip -2\baselineskip plus -1 fil
\begin{IEEEbiography}[{\includegraphics[width=1in,height=1.25in,clip,keepaspectratio]{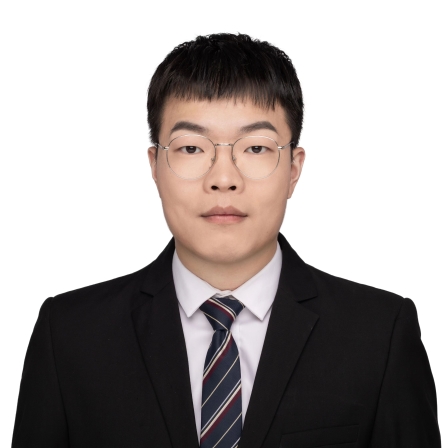}}] {Jianyu Niu} is a Research Assistant Professor in the Research Institute of Trustworthy Autonomous Systems at Southern University of Science and Technology. He received his Ph.D. degree from the School of Engineering, University of British Columbia, Kelowna, Canada, in 2021. His research interests focus on distributed systems, blockchain, and confidential computing.
\end{IEEEbiography}

\vskip -2\baselineskip plus -1 fil
\begin{IEEEbiography}[{\includegraphics[width=1in,height=1.25in,clip,keepaspectratio]{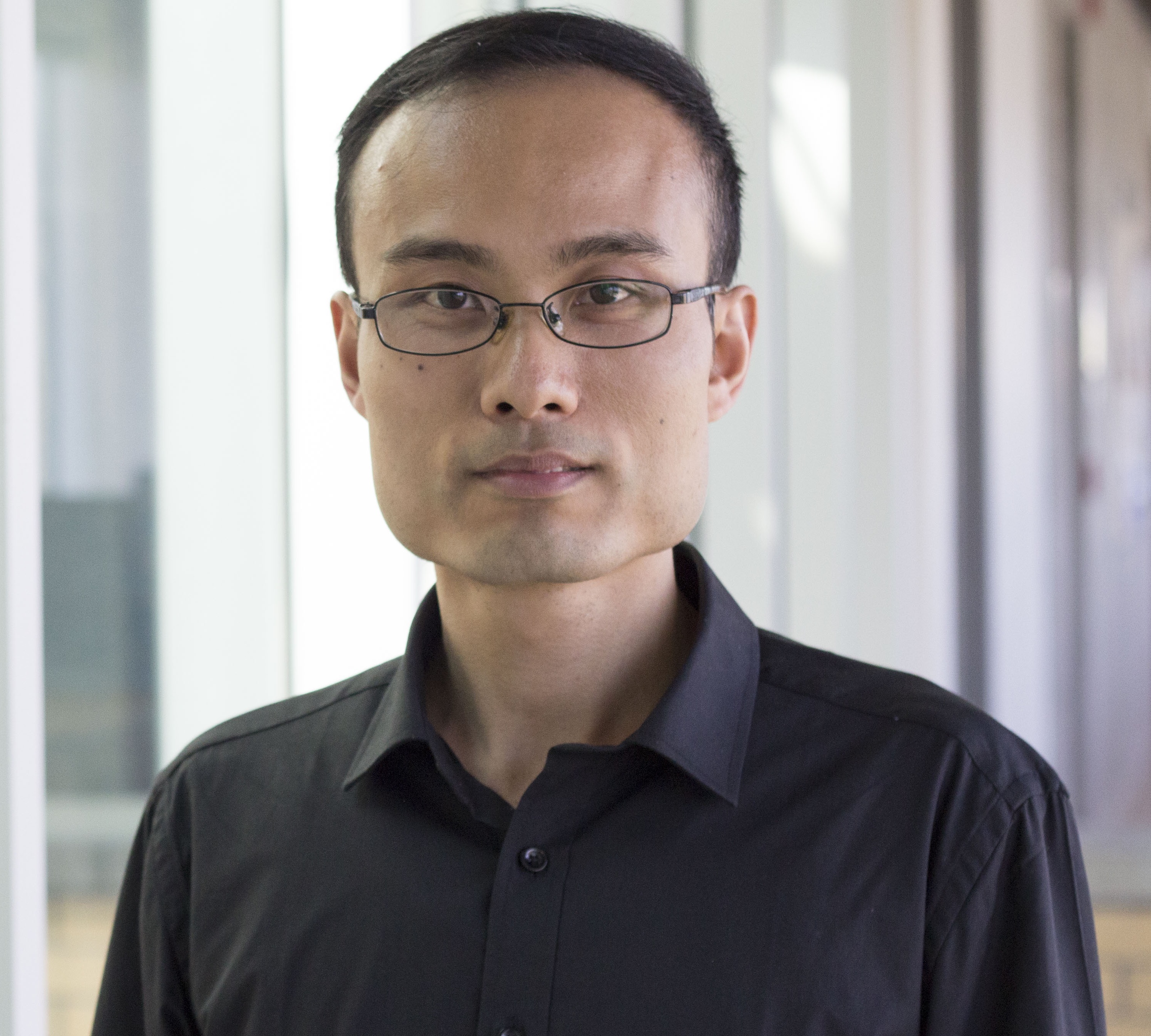}}]{Chen Feng}
    received the B.Eng. degree from Shanghai Jiao Tong University, China, in 2006, and the M.A.Sc. and Ph.D. degrees from The University of Toronto, Canada, in 2009 and 2014, respectively. From 2014 to 2015, he was a Postdoctoral Fellow with Boston University, USA, and EPFL, Switzerland.

    He joined the School of Engineering, University of British Columbia (Okanagan Campus), Kelowna, Canada, in July 2015, where he is currently an Associate Professor. He is a co-cluster lead of Blockchain@UBC and Principal’s Research Chair in Blockchain. He is interested in adapting new ideas and tools from information theory, coding theory, stochastic processes, and optimization to design better communication networks, with a particular emphasis on blockchain technology.
\end{IEEEbiography}
    
\vskip -2\baselineskip plus -1 fil
\begin{IEEEbiography}[{\includegraphics[width=1in,height=1.25in,clip,keepaspectratio]{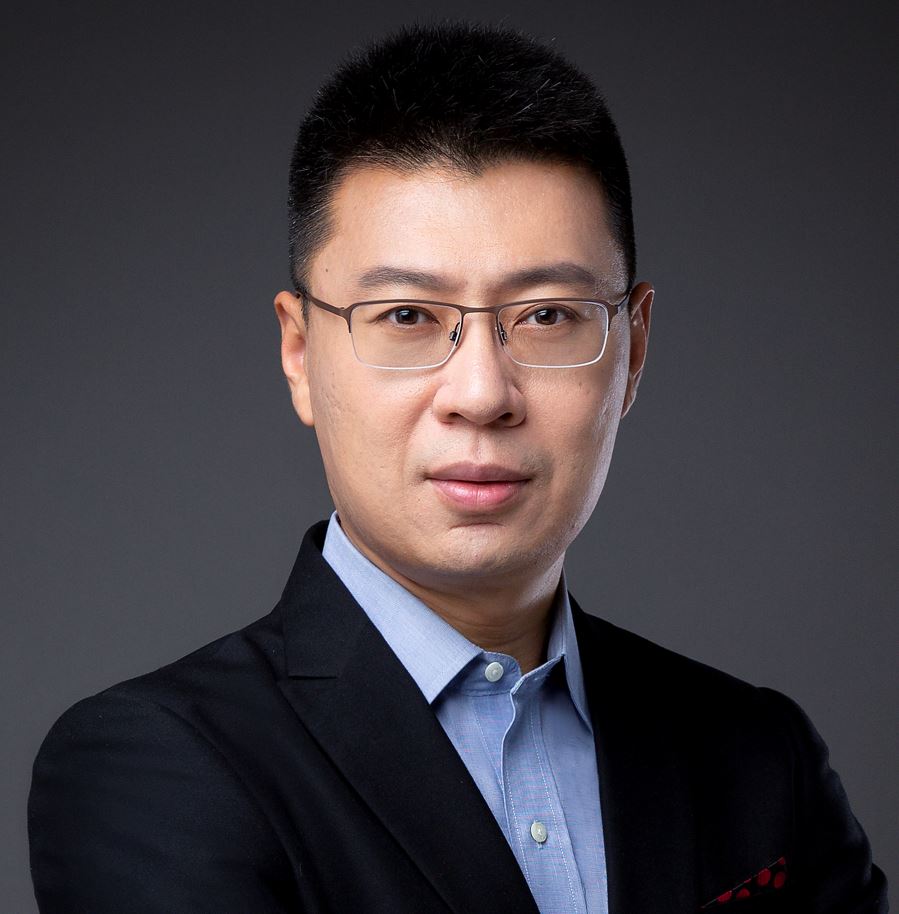}}]{Yinqian Zhang} is a Professor in the Department of Computer Science and Engineering at Southern University of Science and Technology. His research interest lies in system security, including side channels, trusted and confidential computing, and cloud security. 
\end{IEEEbiography}

\end{document}

%% file: macro.tex
\newcommand{\bheading}[1]{{\vspace{4pt}\noindent{\textbf{#1}}}}
\newcommand{\iheading}[1]{{\vspace{1pt}\noindent{\textit{#1}}}}

\newcounter{note}[section]

\newcommand{\secref}[1]{\mbox{Sec.~\ref{#1}}\xspace}

\newcommand{\figref}[1]{\mbox{Fig.~\ref{#1}}}
\newcommand{\tabref}[1]{\mbox{Table~\ref{#1}}}

\newcommand{\ignore}[1]{}
\newcommand{\ssecref}[1]{\mbox{\S\ref{#1}}\xspace}

\newcommand{\cs}{$\texttt{cS}$\xspace}
\newcommand{\lh}{$\texttt{$l_h$}$\xspace}
\newcommand{\la}{$\texttt{$l_a$}$\xspace}
\newcommand{\leader}{$\texttt{$L$}$\xspace}

\newcommand{\metricOne}{chain growth\xspace}
\newcommand{\metricOneUp}{Chain growth\xspace}
\newcommand{\metricTwo}{commitment rate\xspace}
\newcommand{\metricTwoUp}{Commitment rate\xspace}

\newcommand{\Bh}{$\texttt{$B_h$}$\xspace}
\newcommand{\C}{$\texttt{$C$}$\xspace}
\newcommand{\T}{$\texttt{$T$}$\xspace}

\newcommand{\CHS}{CHS\xspace}
\newcommand{\TCHS}{2CHS\xspace}
\newcommand{\FHS}{FHS\xspace}
\newcommand{\Streamlet}{Streamlet\xspace}

\newcommand{\ie}{\textit{i.e.}\xspace}
\newcommand{\eg}{\textit{e.g.}\xspace}
\newcommand{\cf}{\textit{cf.}\xspace}

\newcounter{packednmbr}

\newenvironment{packeditemize}{
\begin{list}{$\bullet$}{
\setlength{\labelwidth}{0pt}
\setlength{\itemsep}{2pt}
\setlength{\leftmargin}{\labelwidth}
\addtolength{\leftmargin}{\labelsep}
\setlength{\parindent}{0pt}
\setlength{\listparindent}{\parindent}
\setlength{\parsep}{1pt}
\setlength{\topsep}{1pt}}}{\end{list}}

\newcounter{lessoncount}